\documentclass[12pt]{article}
\usepackage{amsmath}
\usepackage{geometry}
\usepackage{multirow}
\usepackage{setspace}
\usepackage{fourier}
\usepackage{graphicx,grffile}
\usepackage[caption=false,lofdepth,lotdepth]{subfig}
\usepackage{dcolumn}
\usepackage{float}

\usepackage[utf8]{inputenc}
\usepackage{amssymb}

\graphicspath{    {converted_graphics/}    {/}}
\begin{document}

\title{Theory and simulation of shock waves: Entropy production and energy conversion}


\author{Bj\o rn Hafskjold,$^{\ast}$\textit{$^{a}$} Dick Bedeaux,\textit{$^{a}$} Øivind Wilhelmsen,\textit{$^{b}$} and Signe Kjelstrup\textit{$^{a}$}}
\maketitle
\footnotetext{\textit{$^{a}$~PoreLab, Department of Chemistry, Norwegian University of Science and Technology - NTNU, Trondheim, Norway.}}
\footnotetext{\textit{$^{b}$~PoreLab, SINTEF Energy Research, Trondheim, Norway. }}

\abstract


We have considered a shock wave as a surface of discontinuity and computed the entropy production using non-equilibrium thermodynamics for surfaces.
The results from this method, which we call the "Gibbs excess method" (GEM), were compared with results from three alternative methods, all based on the entropy balance in the shock front region, but with different assumptions about local equilibrium.
Non-equilibrium molecular dynamics (NEMD) simulations were used to simulate a thermal blast in a one-component gas consisting of particles interacting with the Lennard-Jones/spline potential.
This provided data for the theoretical analysis.
Two cases were studied, a weak shock with Mach number $M \approx 2$ and a strong shock with $M \approx 6$ and with a Prandtl number of the gas $Pr \approx 1.4$ in both cases.
The four theoretical methods gave consistent results for the time-dependent surface excess entropy production for both Mach numbers.
The internal energy was found to deviate only slightly from equilibrium values in the shock front.
The pressure profile was found to be consistent with the Navier-Stokes equations.
The entropy production in the weak and strong shocks were approximately proportional to the square of the Mach number and decayed with time at approximately the same relative rate.
In both cases, some 97 \% of the total entropy production in the gas occurred in the shock wave.
The GEM showed that most of the shock's kinetic energy was converted reversibly into enthalpy and entropy, and a small amount was dissipated as produced entropy.
The shock waves traveled at almost constant speed and we found that the overpressure determined from NEMD simulations agreed well with the Rankine-Hugoniot conditions for steady-state shocks.

\section{Introduction}
\label{introduction}

The amount of energy carried by a shock wave is considerable, and the wave travels at supersonic speed.
Shock waves produced from explosions, rapid phase transitions, sudden release of pressurized gas, or other blasts, are highly irreversible phenomena. 
Shock waves are therefore both interesting and challenging phenomena to understand and quantify.
Several laboratory and large-scale field experiments have been carried out to determine the impact of blast waves as function of explosion type and strength, distance from the blast, and topology of the surroundings \cite{ning2015shock}.

The basic equations describing the conditions for shock-waves in one dimension were developed in the late 19th century by Rankine and Hugoniot \cite{rankine1870, hugoniot1887}.
In these early studies, the shock wave was considered as a surface of discontinuity with conservation of mass, momentum, and energy.
The shock's properties were described in terms of the differences between the upstream and downstream properties of the bulk fluids (see \textit{e.g.} Hirschfelder et al. \cite{hirschfelder1954}, Hoover and Hoover \cite{hoover2015}, and Uribe \cite{uribe2011}).
The Rankine-Hugoniot relations give a macroscopic description of the state variables in front of, and behind, the wave front, but not the details of \textit{e.g.} energy dissipation at the front.
Application of hydrodynamic theories in the early 20th century gave more details of the shock-wave front, such as its thickness \cite{jouguet1910,becker1922}.
The developments of kinetic theories at about the same time supported and examined the limitations of  the Navier-Stokes equations as applied to shock waves \cite{becker1922}.
It was found that the thickness of the shock-wave front given by the Navier-Stokes equations was too small compared with experiments and improved theories \cite{bird1967}.
Questions were also raised about the consistency between the entropy profile showing a peak at the front position and the second law of thermodynamics \cite{morduchow1949}.
The interest in blast-wave theory and experiments was high during and after the second world war, which led to significant progress in the understanding of shock waves \cite{Friedlander1946, Taylor1950, Freiwald1972}.
The more recent progress in hydrodynamic theory \cite{margolin2020}, kinetic theory \cite{mott1951,al1997generalized}, and extended thermodynamics \cite{taniguchi2014,arima2012}  has given substantial new insight into many properties of shock waves (see \textit{e.g.} Section 9 in Garc\'ia-Col\'in et. al \cite{garcia2008} for a good review).

The development of Direct Simulation Monte Carlo (DSMC) and molecular dynamics (MD) simulations provided the necessary link between experiments and theories.
Klimenko and Dremin \cite{klimenko1978}, Hoover \cite{hoover1979}, Holian \cite{holian1980}, Salomons \cite{salomons1992}, and their coworkers were pioneers with the aims to clarify some of the puzzles in shock-wave theory and demonstrate the applicability of computer simulations.
For instance, the exact thickness of a shock wave had been estimated \cite{mott1951}, but was not exactly known until simulations produced accurate results \cite{hoover1979, salomons1992}.
With the increase in compute power, very detailed analyses can now be made \cite{salomons1992, holian1993, holian1995, uribe2018}.
The status is that several theories work for shocks with Mach numbers up to approximately 2 (which may be defined as "weak shocks") whereas theories for "strong shocks" (with higher Mach numbers) are still being developed.

Despite the fact that shock-wave propagation is an irreversible process, few papers have been concerned with the energy dissipation and entropy production in shock waves.
Brinkley and Kirkwood presented a theory of non-steady shock waves in 1947 which included the concept of energy dissipation and wave speed retardation \cite{brinkley1947}.
At about the same time, Tolman and Fine published a comprehensive paper on entropy production in irreversible processes, including shock waves \cite{tolman1948}.
It has been shown that kinetic energy is not equipartitioned in the shock-wave front \cite{holian2010}, which is a good reason to question the local-equilibrium assumption made in non-equilibrium thermodynamics (NET).
Nevertheless, this assumption was adopted by Velasco and Uribe, who used the Gibbs equation in the normal way for bulk fluids to obtain the entropy production in the shock front \cite{velasco2019}.
By introducing empirical temperature dependencies of the viscosity in combination with the Navier-Stokes-Fourier relations, they got good agreements with results from DSMC for Mach numbers between 1.55 and 9.
There are, however, remaining questions, such as exactly how the kinetic and compression energy carried by a shock wave is dissipated or converted to other forms, in particular when the wave hits an obstacle or a body.
Such questions are important for studying impact of detonations \cite{kinney2013}, in material science \cite{Zhao2017}, formation and collapse of bubbles \cite{Pecha2000}, and traumatic brain injuries from improvised explosive devices (IEDs) \cite{ning2015shock}, to mention a few.

All approaches to shock-wave analyses use conservation of mass, momentum, and energy (see \textit{e.g.} Landau and Lifshitz, ref), shown here for a plane shock wave moving with constant velocity in $x$-direction in a one-component, single-phase fluid:
\begin{align}
\rho (v-v^\text{s})=& c_1 \label{eqn:cons1} \ \ \ \text{(mass)} \\
\textsf{P}_{xx}+\rho (v-v^\text{s})^2=&c_2 \label{eqn:cons2} \ \ \ \text{(momentum)} \\
\rho (v-v^\text{s}) \left [u+\frac{1}{2}(v-v^\text{s})^2 \right ] +\textsf{P}_{xx} (v-v^\text{s}) +J'_{q,x}=&c_3 \ \ \ \text{(energy)} 
\label{eqn:cons3}
\end{align}
where $\rho$ is the mass density, $v$ and $v^\text{s}$ are the streaming velocity and the shock-wave velocity, respectively, in $x$-direction in the stationary frame of reference, $\textsf{P}_{xx}$ the pressure tensor component normal to the shock-wave front, $u$ the internal energy per unit mass, and $J'_{q,x}$ is the measurable heat flux in $x$-direction.
For simplicity, we have not included gravity in these equations.
Under these conditions, the $c_i$'s are constants.
In the classical treatment of shock waves in Newtonian fluids, Navier-Stokes and Fourier constitutive relations are introduced into the conservation laws (in addition to an equation of state) \cite{salomons1992}, \textit{viz.}
\begin{align}
\textsf{P}_{xx}=& p-\left ( \frac{4}{3}\eta_\text{S} + \eta_\text{B} \right ) \frac{\partial v}{\partial x} \label{eqn:const1} \\
\textsf{P}_{yy}=\textsf{P}_{zz}=& p +\left ( \frac{2}{3}\eta_\text{S} - \eta_\text{B} \right ) \frac{\partial v}{\partial x} \label{eqn:const2} \\
J'_q=& -\lambda \frac{dT}{dx} \label{eqn:const3}
\end{align}%
where $\textsf{P}_{yy} = \textsf{P}_{zz}$ is the pressure parallel with the shock front, $p$ is the equilibrium pressure as given by the equation of state at the local conditions, $\eta_\text{S}$ and $\eta_\text{B}$ are the shear and bulk viscosities, respectively, $\lambda$ the thermal conductivity, and $T$ is the temperature.
All the quantities in Eqs. \eqref{eqn:const1} - \eqref{eqn:const3} are in general functions of position $x$ and time $t$.
In the framework of Navier-Stokes, it follows from Eqs. \eqref{eqn:const1} and \eqref{eqn:const2} that if the viscous terms are small, the diagonal components of the pressure tensor are approximately equal to the equilibrium pressure.

A shock wave may be characterized by a sharp front with significant changes in density and pressure over such a short distance that, at a macroscopic scale, it can be considered to be a discontinuity in the system's characteristic properties \cite{hirschfelder1954}. 
This is not unlike the case of a regular surface, \textit{e. g.} a liquid/vapor surface. 
NET for surfaces has been developed by Kjelstrup and Bedeaux \cite{kjelstrup2020} and we recently reported that NET can successfully be applied to a weak shock wave (Mach number 2.1) \cite{hafskjold2020}.
In the present work, we develop the method further, analyze its basis in more detail, and use the results to describe the excess entropy production and energy conversions on the surface.
Furthermore, we find a flux-force relation for mass transfer across the shock front in the surface description.
The work presented here is a new approach to assess the properties of shock waves.
This work includes data for a significantly stronger shock wave than the one we considered in our previous analysis \cite{hafskjold2020}.

We have used four different methods to determine the surface excess entropy production.
The methods are based on different assumptions and the consistency between the methods are used as a criterion for the validity of their underlying assumptions.
The concept of local equilibrium must be defined in different ways depending on the context of the methods we apply.
All four methods are based on the balance equation for entropy across the shock-wave front.
In the "bulk balance method" (BBM) and the linear irreversible thermodynamics (LIT) method we integrate the local entropy production over the shock wave thickness.
The BBM uses the entropy balance directly whereas the LIT involve the Gibbs equation as well.
The "surface balance method" (SBM) models the wave front as a surface of discontinuity and considers the entropy balance on the surface.
In the "Gibbs excess method" (GEM), we use the Gibbs equation for surface variables \cite{kjelstrup2020, deGroot1962} and derive a new, more detailed expression for entropy production and a new tool to analyze the energy conversions in a shock wave.
The basic assumption in GEM is that this Gibbs equation for surface excess variables remains valid also when the system at large is out of equilibrium, as suggested by Bedeaux, Albano and Mazur in their construction of NET for heterogeneous systems \cite{Bedeaux1976, Albano1987}.
Energy dissipation and conversion can be determined from this analysis, which leads to new information about energy conversion at the shock front.
We show that a particularly simple constitutive relation can be used for shocks at, or close to, steady state.
This relation has not been used in earlier work.
At present, we restrict the discussion to one-component fluids.

The analysis starts with the balance equation for entropy in Section \ref{balance}.
The BBM and LIT are based on this balance.
In Section \ref{theory} we establish the framework for the surface description of shock waves.
This includes the definition of the Gibbs equimolar surface and reformulation of the balance equations and conservation laws in the surface description.
The SBM is based on this surface description.
The new relations based on NET are derived in Section \ref{surface method} where we specify the conditions and assumptions made.
The key result in this section is an expression for the surface excess entropy production.
In order to quantify the theoretical results, we have done non-equilibrium molecular dynamics (NEMD) simulations of two shock waves, one weak and one strong (Mach number $M \approx 2$ and $M \approx 6$, respectively).
Section \ref{simulations} includes a description of the model system used in the NEMD simulations and how the simulations were carried out. 
We show in Section \ref{results} that NEMD is well suited to analyze the theories and provide unbiased data. 
The combination of NET theory and NEMD data lead to a new insight into the various contributions to the entropy production in a shock wave.
Section \ref{results} also includes  descriptions of how the NEMD data were used to examine the assumptions mentioned above (Sections \ref{pressureprofiles} and \ref{velocityprofiles}).
Finally, we summarize the conclusions from this work in Section \ref{conclusions}.

\section{Basic entropy balance and the meanings of "local"}
\label{balance}

\begin{figure}[tbp]
\centering
\includegraphics[trim=0 10 0 30, clip, scale=0.4]{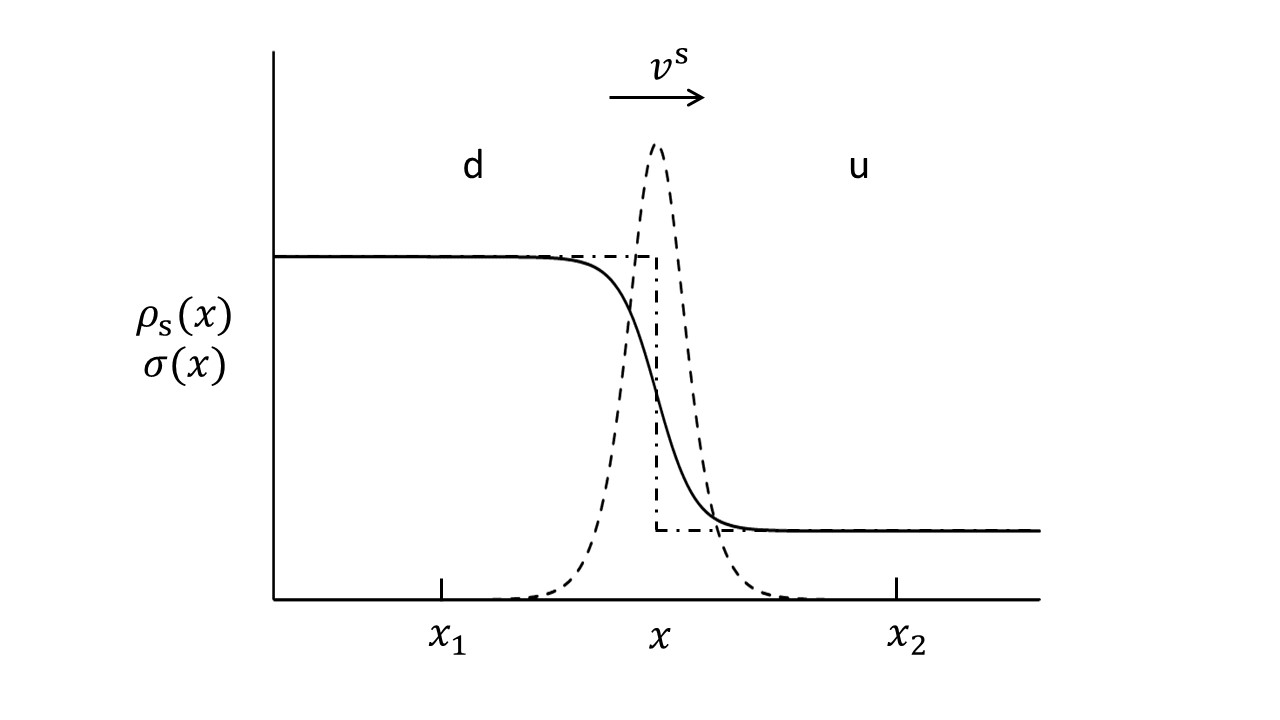}
\caption{Illustration of the entropy density, $\rho_\text{s}$, (solid line) and the entropy production, $\sigma_\text{s}$, (dashed line) around a shockwave moving from left to right with velocity $v^\text{s}$. The dash-dot-line is a macroscopic illustration of the shockwave with discontinuous fluid properties.}
\label{fig:sketch}
\end{figure}

In addition to the conservation laws for mass, momentum, and energy, Eqs. \eqref{eqn:cons1}-\eqref{eqn:cons3}, we consider the entropy over an infinitesimal control volume in one dimension:
\begin{equation}
\frac{\partial}{\partial t} \rho_\text{s}(x,t)+\frac{\partial }{\partial x}J_\text{s}(x,t)=\sigma_\text{s}(x,t)
\label{eqn:B.1}
\end{equation}
where $\rho_\text{s}$, $J_\text{s}$, and $\sigma^\text{s}$ are the density, flux, and production of entropy, respectively.
Profiles of the entropy density and entropy production, centered around a shock wave front, are sketched in Figure \ref{fig:sketch}.
The entropy flux is a combination of heat conduction and entropy transported with the fluid flow,
\begin{equation}
J_\text{s}(x,t) = \frac{J_\text{q}^ \prime (x,t)}{T(x,t)} + \rho_\text{s}(x,t) v(x,t)
\label{eqn:B}
\end{equation}
where $v$ is the local fluid velocity.
Whereas the heat conduction is independent of the frame of reference, the transported entropy is not, and is here given in the stationary frame of reference.

In Eqs. \eqref{eqn:B.1} and \eqref{eqn:B} we have used the term "local" in the meaning "infinitesimally small domains in space and time".
In the following, we shall use "local" in two different meanings, depending on whether we consider the microscopic picture and the wave front as a spatial domain with continuously varying properties, or the macroscopic picture and the wave front as a surface.
"Local" in the former context means a small control volume, which in the numerical work is determined by the thickness of each layer in the NEMD simulations, typically about 30 molecular diameters or slightly more than the mean free path in the equilibrium gas ahead of the shock.
"Local" in the latter sense means "on the surface", \textit{i.e.} as given by the surface excess properties.
The surface has no extension in $x$-direction.
In both contexts, "local" in time means a time interval determined by the sampling time in the numerical simulations.

We shall now consider the four routes to the surface excess entropy production in the shock wave based on Eq. \eqref{eqn:B.1}.
The basic route is a direct integration of Eq. \eqref{eqn:B.1}, which we call the "bulk balance method" (BBM).
The results from BBM are discussed in Section \ref{entropybalance}.
In two other routes we consider the shock wave front as a surface and apply NET for surfaces where we make use of the equilibrium bulk properties at both sides of the wave.
The fourth method is the classic linear irreversible thermodynamics (LIT) method based on the assumption of local equilibrium and the Gibbs equation for bulk systems \cite{deGroot1962}.
When applied to each local control volume in the system, this leads in the present context to 
\begin{equation}
\sigma_\text{s}(x)= J_q^\prime \frac{\partial}{\partial x} \left ( \frac{1}{T} \right ) - \frac{1}{T} \Pi_{xx} \frac{\partial v}{\partial x}
\label{eqn:B.1a}
\end{equation}
where $\Pi_{xx}$ is the $x$-component of the viscous pressure tensor.
The LIT method was recently used by Velasco and Uribe in an analysis of shock waves \cite{velasco2019}.

\section{The shockwave as a surface}

\label{theory}

A typical surface has a thickness that is small compared to the thickness of the adjacent homogeneous phases, and the surface appears to be two-dimensional.
In reality, a shock wave is several molecular diameters thick, depending on the system's thermodynamic state and speed of the wave \cite{mott1951}, in this study about 100 molecular diameters.
At the macroscopic scale, transport of heat and matter across a surface will give rise to discontinuities in intensive variables like temperature and chemical potential. 
The fluxes and forces may also become discontinuous. 
Our derivation of NET for shock waves builds on Gibbs' definition of a surface \cite{Gibbs1961}, and the assumption first made by Bedeaux, Albano and Mazur \cite{Bedeaux1976, Albano1987}, \textit{viz.} that thermodynamic relations between surface variables remain valid locally, also when the system overall is out of equilibrium.
This assumption means that we define the interface as a separate and autonomous thermodynamic system \cite{hafskjold2020}.
The surface is assumed to possess a temperature, chemical potential, and other thermodynamic variables of its own.
The assumption may seem drastic because the shock-wave front is in a non-equilibrium state without a corresponding equilibrium state (in contrast to the case for \textit{e.g.} a liquid-vapor surface).
We shall therefore examine this assumption in detail in Section \ref{Gibbs}.

Thermodynamic properties of surfaces are well defined using Gibbs' \textit{surface excess densities} of mass, entropy and energy \cite{Gibbs1961}.
Following the systematic procedure given by Albano \textit{et al.} \cite{Albano1987}, we first derive the entropy production on the surface.

\subsection{The Gibbs surface}
\label{gibbsequalarea}

Gibbs defined the \textit{equimolar surface} as \textquotedblleft a geometrical plane, going through points in the interfacial region, similarly situated with respect to conditions of adjacent matter\textquotedblright \cite{Gibbs1961}. 
Many different positions can be chosen for a plane of this type. 
If the density of a quantity "a", $\rho_\text{a}$, varies in $x$-direction according to $\rho_\text{a}(x,y,z)$, the \textit{excess density} $\rho_\text{a}^{\text{s}}$ is
\begin{equation}
\rho_\text{a}^{\text{s}}(y,z)=\int_{x_1}^{x_2}\left[ \rho_\text{a}(x,y,z)-\rho_\text{a}^{\text{d}}(x,y,z)\Theta
(\ell -x)-\rho_\text{a}^{\text{u}}(x,y,z)\Theta (x-\ell )\right] dx  
\label{eqn:G.1}
\end{equation}
Here, $x_1$ and $x_2$ are positions in the bulk phases and $\ell $ is the position of the equimolar surface.
The superscripts "d" and "u" indicate a function used to extrapolate $\rho_\text{a}(x,y,z)$ from the bulk values on the left (downstream) and right (upstream) side, respectively, of the wave as illustrated by the dash-dot-line in Figure \ref{fig:sketch}.
The figure also illustrates the positions $x_1$ and $x_2$.
The Heaviside step function, $\Theta$, is by definition unity when the argument is positive and zero when the argument is negative. 
Note that whereas the bulk density is per unit volume, the excess density is per unit area.

The excess density is in general a function of the position $(y,z)$ along the surface. We shall, however, only consider the case of constant properties in the $y,z$ plane, and these coordinates will be omitted from here on.
Moreover, the cross-sectional area perpendicular to the $x$-direction is independent of $x$.
Examples of $\rho_\text{a}$\ considered below are the mass, momentum, energy, and entropy densities. 

\begin{figure}[tbp]
\includegraphics[trim=0 50 0 30, clip, scale=0.5]{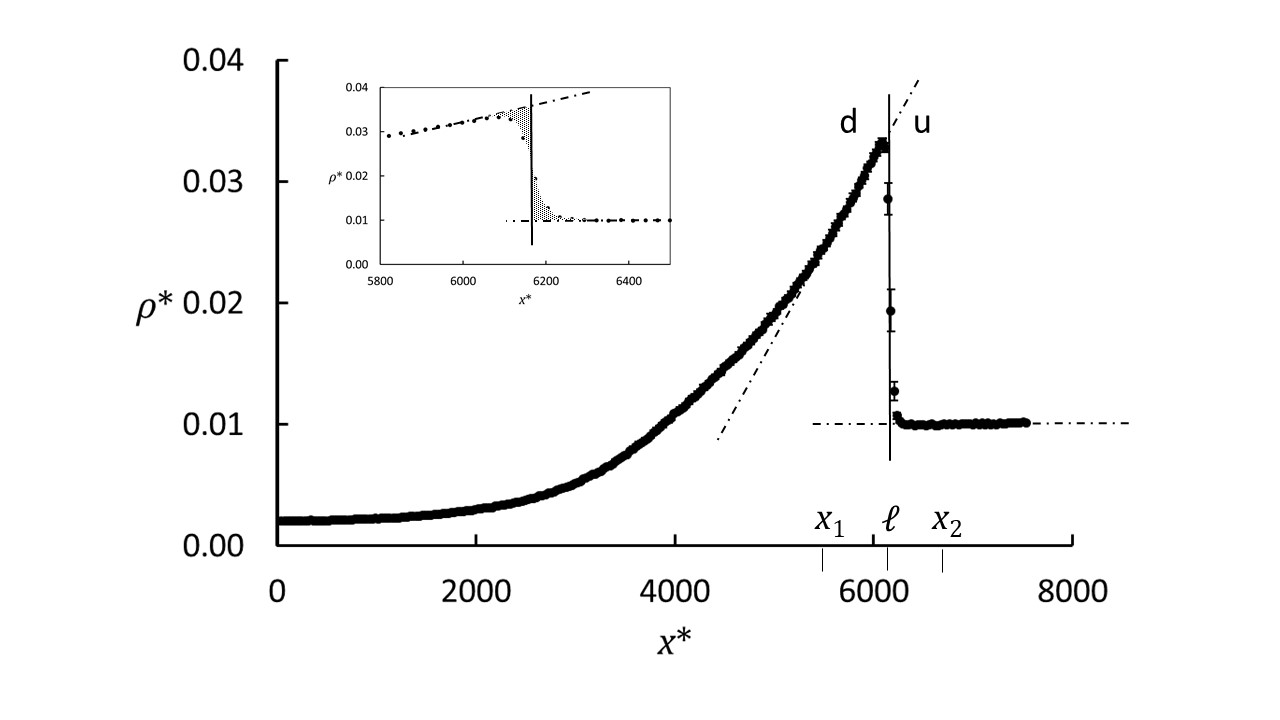}
\caption{Gibbs' equal-area construction for determination of shock-front position, $\ell$, by requiring that the surface excess mass density $\rho ^\text{s}=0$ (the two shaded areas in the insert are equal). 
The black circles are results from the NEMD simulations for the strong shock as described in Section \ref{simulations}. 
The vertical line shows the position of the equimolar surface and the dashed-dot lines are least-squares fit to the bulk data.
The uncertainties in the NEMD data, determined as three standard errors, are shown as vertical bars.}
\label{fig:gibbsequalarea}
\end{figure}

All surface excess properties can be given by integrals like Eq. \eqref{eqn:G.1}.
We shall first consider the mass density $\rho$.
Requiring the excess molar density to be zero defines the \textit{equimolar surface}, which we shall use to define the position $\ell$ of the surface.\footnote{In the present one-component case, this is equivalent to use the excess mass density.}
It is chosen such that the surplus of matter on one side of the surface is equal to the deficiency on the other side. 
The shockwave position is a function of time $t$, $\ell (t)$.
The velocity of the surface is given by
\begin{equation}
v^{\text{s}}(t)=\frac{d\ell (t)}{dt}  
\label{eqn:B.3}
\end{equation}
in the stationary frame of reference.
If $v^{\text{s}}(t)$ is independent of $t$, the surface moves at steady state.

Like the mass density, other excess variables are given per unit area of the surface.
They describe the surface and how it \textit{differs} from the adjacent
homogeneous phases.
In particular, the surface excess mass density $\rho^\text{s}$ of the equimolar surface is zero.
With the surface location so defined, other surface excess variables will in general be non-zero.
Within reasonable limits, one may shift the positions $x_1$ and $x_2$ without changing the extrapolated values of interest. 
In this sense, the precise locations of $x_1$ and $x_2$ are not important for the value of the excess property as long as they are in the bulk phases near the surface as illustrated in Figure \ref{fig:gibbsequalarea}.


\subsection{Balance equations in the surface frame of reference}

\label{2.2}

The density, flux, and production of a surface property "a", $\rho_\text{a}$, $J_\text{a}$, and $\sigma_\text{a}$, respectively, obey dynamic balance equations.
At the macroscopic scale, we may write $\rho_\text{a}$ in the form \cite{kjelstrup2020}
\begin{equation}
\rho_\text{a}^\prime(x,t)=\rho_\text{a}^{\text{d}}(x,t)\Theta (\ell (t)-x)+\rho_\text{a}^{\text{s}}(t)\delta (x-\ell (t))+\rho_\text{a}^{\text{u}}(x,t)\Theta (x-\ell (t)).  \label{eqn:B.2}
\end{equation}
The limit of the Gaussian-type distribution function shown as the dashed line in Figure \ref{fig:sketch} is the Dirac delta function in the surface description.
Fluxes $J_\text{a}$\ and source terms $\sigma_\text{a}$\ are given by similar expressions.

By substituting Eq. \eqref{eqn:B.2} into Eq. \eqref{eqn:B.1} and using Eq. \eqref{eqn:B.3} we obtain \cite{Bedeaux1976}
\begin{align}
&\left[ \frac{\partial}{\partial t} \rho_\text{a}^\text{d}(x,t) +\frac{\partial }{
\partial x}J_\text{a}^\text{d}(x,t)-\sigma_\text{a}^\text{d}(x,t)\right] \Theta
(\ell (t)-x) \nonumber \\
&+\left[ \frac{\partial}{\partial t} \rho_\text{a}^\text{u}(x,t) +\frac{\partial }{
\partial x}J_\text{a}^\text{u}(x,t)-\sigma_\text{a}^\text{u}(x,t)\right] \Theta
(x-\ell (t)) \nonumber \\
&+\left[ \frac{d}{dt}\rho_\text{a}^\text{s}(t) +J_\text{a}^\text{u}(\ell,t)-v^\text{s}(t)\rho_\text{a}^\text{u}(\ell,t)-J_\text{a}^\text{d}(\ell,t)+v^\text{s}(t)\rho_\text{a}^\text{d}(\ell,t)-\sigma_\text{a}^\text{s}(t)\right] \times \delta (x-\ell (t)) \nonumber \\
&+\left[ J_\text{a}^\text{s}(t)-v^\text{s}(t)d_\text{a}^\text{s}(t)\right] \frac{\partial}{\partial x} \delta (x-\ell (t)) \nonumber \\
&=0
\label{eqn:B.4}
\end{align}
where it is understood that $J_\text{a}^{\text{u}}(\ell,t)$, $J_\text{a}^{\text{d}}(\ell,t)$, $\rho_\text{a}^\text{u}(\ell,t)$, and $\rho_\text{a}^\text{d}(\ell,t)$ in the third line are the extrapolated values (to the surface position) of the respective quantities.
In order for Eq. \eqref{eqn:B.4} to be correct, the sum of all terms inside each of the square brackets have to be zero. 
The first two brackets give equations for the bulk phases in the macroscopic description. 
For the surface we obtain from the third bracket
\begin{equation}
\frac{d}{dt}\rho_\text{a}^{\text{s}}(t)+\left[ J_\text{a}(t)-v^{\text{s}}(t)\rho_\text{a}(t)\right]
_{-}=\sigma_\text{a}^{\text{s}}(t) 
\label{eqn:B.5}
\end{equation}
where we have used the notation 
\begin{equation}
\left [ J_\text{a}(t)-v^{\text{s}}(t)\rho_\text{a}(t) \right ] _- \equiv J_\text{a}^{\text{u}}(t)-v^{\text{s}}(t)\rho_\text{a}^{\text{u}}(t)-J_\text{a}^\text{d}(t)+v^\text{s}(t)\rho_\text{a}^{\text{d}}(t)  
\label{eqn:B.6}
\end{equation}
for the difference across the surface. 
Equation \eqref{eqn:B.5} shows that the accumulation of the property "a" on the surface is due to the difference in the flux (in the surface frame of reference) in and out of the surface plus the excess production. 
In particular, if we consider the mass density $\rho$, we find
\begin{equation}
\frac{d}{dt}\rho^{\text{s}}(t)+\left[ j \right]
_{-}=0 
\label{eqn:B.5a}
\end{equation}
where
\begin{equation}
j=\rho(v-v^{\text{s}})
\label{eqn:j}
\end{equation}
is the mass flux in the surface frame of reference.
By construction, $\rho^{\text{s}}(t)=0$.
This gives the mass conservation, Eq. \eqref{eqn:cons1}, in the surface description:
\begin{equation}
\left[ j \right]_{-}=0 
\label{eqn:ja}
\end{equation}

The balance equation for the surface excess entropy production follows from Eq. \eqref{eqn:B.5}:
\begin{equation}
\frac{d \rho_\text{s}^{\text{s}}}{dt}+\left[ J_\text{s}-v^{\text{s}}\rho_\text{s}\right] _{-}=\sigma_{\text{s}}^{\text{s}}
\label{eqn:S.1}
\end{equation}%
Here $\rho_\text{s}^{\text{s}}$ is the surface excess entropy density (per unit surface area).
Eq. \eqref{eqn:S.1} is given in terms of entropy fluxes into and out of the surface (in the surface frame of reference), and the excess entropy production, $\sigma_{\text{s}}^{\text{s}}$.
Although Eq. \eqref{eqn:B.1} allows us to determine the entropy production in the entire system, we focus here on the surface for which $x$ and $t$ are related through the temporal position of the surface.
Eq. \eqref{eqn:S.1} is basis for the SBM, one of the two surface methods we will use to determine $\sigma_{\text{s}}^{\text{s}}$  quantitatively in Section \ref{results}.

The fourth term in Eq. \eqref{eqn:B.4} gives
\begin{equation}
J_\text{a}^{\text{s}}(t)-v^{\text{s}}(t)\rho_\text{a}^{\text{s}}(t)=0  \label{eqn:B.7}
\end{equation}
which implies that the excess flux of the quantity "a" in the direction of the shock-wave propagation is equal to zero in the frame of reference that moves with the shock.

%
%
%
%

\subsection{Conservation laws}

We can now apply the general considerations in Section \ref{2.2} to the
conservation of mass, momentum, and energy. 
From the conservation of mass it follows that
\begin{equation}
\frac{\partial \rho }{\partial t}+\frac{\partial (\rho v)}{\partial x}=0
\label{eqn:C.1}
\end{equation}
in the bulk phases.
Eq. \eqref{eqn:ja} describes conservation of mass for the surface.
From conservation of momentum, it follows that
\begin{equation}
\frac{\partial (\rho v)}{\partial t}+\frac{\partial }{\partial x}\left( \textsf{P}_{xx}+\rho vv\right) =0 
\label{eqn:C.3}
\end{equation}%
in the bulk phases and
\begin{equation}
\frac{d (\rho v)^{\text{s}}}{dt}+\left[ \textsf{P}_{xx} + jv \right] _{-}=0 
\label{eqn:C.4}
\end{equation}
for the surface. 
In Eqs. \eqref{eqn:C.3} and \eqref{eqn:C.4}, $\textsf{P}_{xx}=p+\Pi_{xx}$ where $p$ is the thermodynamic pressure and $\Pi_{xx}$ is the $xx$-component of the viscous pressure tensor.
Making the Navier-Stokes assumption, $\Pi_{xx}=-\left ( \frac{4}{3}\eta_\text{S}+\eta_\text{B} \right ) \frac{\partial v}{\partial x}$.

From conservation of energy it follows that
\begin{equation}
\frac{\partial \rho_\text{e}}{\partial t}+\frac{\partial }{\partial x}\left( \rho_\text{e}v+\textsf{P}_{xx}v+J_{q}\right) =0  
\label{eqn:C.5}
\end{equation}%
where the energy density $\rho_\text{e}$ is the sum of internal and kinetic energy density: $\rho_\text{e}=\rho_\text{u}+\rho_\text{k}$, where $\rho_\text{k}=\rho v^2 /2$, and $J_{q}$
is the total heat flux in the barycentric frame of reference, all in the
bulk phases. 
In the one-component system that we consider, $J_{q}=J_{q}^{\prime }$ where $J_{q}^{\prime }$ is the measurable heat flux, which is independent of the frame of reference.
Furthermore,
\begin{equation}
\frac{d \rho_\text{e}^{\text{s}}}{d t}+\left[ \rho_\text{e} (v-v^{\text{s}}) +\textsf{P}_{xx} v+J_{q}^{\prime} \right] _{-}=0 
\label{eqn:C.6}
\end{equation}%
for the surface.
By analogy to the bulk energy density, $\rho_\text{e}^{\text{s}}=\rho_\text{u}^{\text{s}}+\rho_\text{k}^{\text{s}}$ where $\rho_\text{k}^{\text{s}}=(\rho v^2)^\text{s}/2$ for the surface.
The sum of properties in the bracket, $\rho_\text{e} (v-v^\text{s}) +\textsf{P}_{xx} v+J_{q}^\prime$, extrapolated to the surface, is the energy flux in the surface frame of reference.\footnote{At stationary state, this means that the difference between the upstream and downstream sums in the bracket is zero.}

For the excess internal energy density it follows from Eqs. \eqref{eqn:C.4} and \eqref{eqn:C.6} that \footnote{Because the surface speed $v^\text{s}$ is connected to the excess surface mass density by the definition of the surface, the time derivative of the kinetic energy density in Eq. \eqref{eqn:C.6} equals the surface velocity times the time derivative of the momentum density.}

\begin{equation}
\frac{d \rho_\text{u}^{\text{s}}}{dt} +\left[ j \left \{ h + \frac{\Pi_{xx}}{\rho}+\frac{1}{2} \left( v-v^{\text{s}}\right )
^{2} \right \}  +J_{q}^\prime \right]_{-} =0
\label{eqn:C.7}
\end{equation}
where the specific enthalpy is $h=u+p$.
\section{The Gibbs excess method (GEM)}

\label{surface method}

\subsection{The entropy production}
\label{Gibbsequation}

So far, we have three routes to the surface excess entropy production, the BBM and the LIT (the integrals of Eq. \eqref{eqn:B.1} and Eq. \eqref{eqn:B.1a}, respectively, over the surface thickness), and the SBM, Eq. \eqref{eqn:S.1}.
We now proceed to find a third route using the Gibbs equation for the surface. 
We shall see that $\sigma_{\text{s}}^{\text{s}}$ can to a very good approximation be written as the product of a mass flux and the entropy difference across the surface.
Furthermore, the GEM provides detailed information about the energy conversions in the shock wave.

The integrated form of the Gibbs equation for a surface is \cite{kjelstrup2020},
\begin{equation}
\rho_\text{u}^{\text{s}}=T^{\text{s}}\rho_\text{s}^{\text{s}}+\gamma + \rho^{\text{s}} \mu^{\text{s}}
\label{eqn:S.2}
\end{equation}
where $T^\text{s}$ is the surface temperature,\footnote{$T^\text{s}$ has a value that in general differs from the temperatures in the bulk phases.
} $\gamma $\ is defined by $\gamma=(\partial U^\text{s}/\partial \Omega)_{\{ S^\text{s},N^\text{s}\} }$ where $U^\text{s}$, $S^\text{s}$ and $N^\text{s}$ are the surface excess internal energy, entropy, and number of particles, respectively,\footnote{Note that the upper-case symbols mean extensive properties of the surface.} and $\mu^{\text{s}}$ is the specific Gibbs energy of the surface.
When Eq. \eqref{eqn:S.2} is combined with the Gibbs-Duhem equation,
\begin{equation}
\rho_\text{s}^{\text{s}}dT^{\text{s}}+d\gamma + \rho^{\text{s}} d \mu^{\text{s}}=0,
\end{equation}
we find
\begin{equation}
d \rho_\text{u}^{\text{s}}=T^{\text{s}}d \rho_\text{s}^{\text{s}}+ \mu^{\text{s}} d \rho^{\text{s}} = T^{\text{s}}d \rho_\text{s}^{\text{s}} 
\label{eqn:Gibbslocal}
\end{equation}
The second equality is due to the fact that the surface excess density $\rho^{\text{s}}$ is zero by construction.
Eq. \eqref{eqn:Gibbslocal} is the statement of local equilibrium in the surface description.
The statement implies that the surface excess properties are related in a way that can be used to assess the entropy production.
The time derivative of the excess entropy density is: 
\begin{equation}
\frac{d\rho_\text{s}^{\text{s}}}{dt}=\frac{1}{T^{\text{s}}}\frac{d \rho_\text{u}^{\text{s}}}{dt}
\label{eqn:S.3}
\end{equation}
where $\rho_\text{s}^{\text{s}}$ and $\rho_\text{u}^{\text{s}}$ are determined from Eq. \eqref{eqn:G.1} using $\ell$ from the equimolar surface.
Eq. \eqref{eqn:S.3} then gives the surface temperature $T^{\text{s}}$.

By introducing Eq. \eqref{eqn:C.7} into Eq. \eqref{eqn:S.3}, and
comparing the result with the entropy balance, Eq. \eqref{eqn:S.1}, we obtain the following expression for the excess entropy production, using the same bracket notation as in Eq. \eqref{eqn:B.6}:
\begin{equation}
\sigma_{\text{s}}^{\text{s}} =[\sigma_q]_- + [ \sigma_j]_-,
\label{eqn:N}
\end{equation}
where 
\begin{equation}
\sigma_q =J_q'\left( \frac{1}{T} - \frac{1}{T^\text{s}}\right)
\label{eqn:Na}
\end{equation}
and
\begin{equation}
\sigma_j = j \left \{ s - \frac{1}{T^{
\text{s}}} \left (  h + \frac{\Pi_{xx}}{\rho} +\frac{1}{2} ( v-v^{\text{s}}) ^{2} \right) \right \}
\label{eqn:Nb}
\end{equation}
where $s$ is the specific entropy.
Eqs. \eqref{eqn:Na} and \eqref{eqn:Nb} contain quantities that are available from the equation of state plus the thermal conductivity and the viscosity.
The results from this method will be compared with results from the other three methods in Section \ref{results}.

The excess entropy production is independent of the frame of reference, but as a property of the surface, it will in general depend on how the surface is defined.
It is in other words invariant under a coordinate transformation.
We may therefore convert all fluxes and conjugate forces from any frame of reference, to the surface frame of reference and back, without changing the entropy production in the different phases, $\sigma_{\text{s}}^{\text{d}}$, $\sigma_{\text{s}}^{\text{s}}$, and $\sigma_{\text{s}}^{\text{u}}$. 

\subsection{Stationary shock front}

\label{front}

If the shock front moves with a constant velocity, all shock-front variables, except for the position of the shock front, are independent of time. 
The conservation equations \eqref{eqn:C.4} and \eqref{eqn:C.7} then reduce to the Rankine-Hugoniot conditions, which in our notation are given by 
\begin{equation}
\left[ \textsf{P}_{xx} + jv \right] _{-}=0 
\end{equation}
\begin{equation}
\left[ j \left\{ h + \frac{\Pi_{xx}}{\rho}+\frac{1}{2} \left( v-v^{\text{s}}\right )
^{2} \right \} +J_{q}^\prime \right] _{-} =0
\label{eqn:C.8}
\end{equation}
Eq. \eqref{eqn:C.8} can be used to eliminate $T^\text{s}$ in Eq. \eqref{eqn:Nb}.
Since the upstream system is at equilibrium, the upstream heat flux equals zero.
The downstream heat flux is close to zero because the temperature gradient just behind the wave front is small (\textit{cf} Figure \ref{fig:temperature}).
Under these conditions we can therefore neglect the contribution to the entropy production from the heat flux (Eq. \eqref{eqn:Na}) and approximate 
\begin{equation}
\sigma_{\text{s}}^{\text{s}} \approx j \left [ s \right] _{-}
\label{eqn:sigmaapprox}
\end{equation}
Eq. \eqref{eqn:sigmaapprox} also follows from Eq. \eqref{eqn:S.1} with $\frac{d \rho_\text{s}^{\text{s}}}{dt}=0$. 
In a transient state, like we have studied in this paper, both $j$ and $s^\text{d}$ vary with time, so Eq. \eqref{eqn:sigmaapprox} will approximate a time-dependent $\sigma_{\text{s}}^{\text{s}}$ even though it is based on a steady-state approximation.

\subsection{Constitutive equations}
\label{constitutive}

Eq. \eqref{eqn:sigmaapprox} gives a particularly simple flux-force relation in the surface description with just one flux ($j$), one force ($\left [ s \right]_{-}$), and one transport coefficient ($L$):
\begin{equation}
j=L \left [ s \right]_{-}
\label{eqn:consta}
\end{equation}
For the sake of completeness, we also include here the corresponding equations from the entropy production, Eq. \eqref{eqn:B.1a}.
Note that the two terms on the right hand side of Eq. \eqref{eqn:B.1a} are of different tensorial order and therefore do not couple, so that the constitutive equations are in this description:
\begin{align}
J_q^\prime=&L_{qq} \frac{\partial}{\partial x} \left ( \frac{1}{T} \right ) = - \lambda \frac{\partial T}{\partial x} \label{eqn:constb}\\
\Pi_{xx} =&  L_{\Pi\Pi} \frac{1}{T} \frac{\partial v}{\partial x} = - \left ( \frac{4}{3}\eta_\text{S} + \eta_\text{B} \right ) \frac{\partial v}{\partial x}
\label{eqn:constc}
\end{align}
where the last terms on the right-hand side of both equations are the Fourier-Navier-Stokes constitutive equations.
Eqs. \eqref{eqn:constb} and \eqref{eqn:constc} are valid for both steady and transient states.

\section{Non-equilibrium molecular dynamics simulations of a blast wave}

\label{simulations} 

NEMD simulations were carried out with a Lennard-Jones/spline (LJ/s) model using an in-house Fortran code.
The model is defined by the pair potential

\begin{equation}
u(r)=
  \begin{cases}
  4\varepsilon\left[ \left( \frac{\sigma}{r}\right)^{12}-\left( \frac{\sigma}{r}\right)^{6}  \right]  & \text{if }r<r_s \\ 
  a(r-r_c)^2+b(r-r_c)^3 & \text{if }r_{s}<r<r_c  \\ 
  0 & \text{if }r>r_c
  \end{cases}
  \label{eqn:ljs}
\end{equation}
where $\sigma$ and $\varepsilon$ are the usual Lennard-Jones potential parameters and $a$ and $b$ are coefficients in the spline function that truncates the potential smoothly between the potential's inflection point at $r_s$ and zero value at $r_c$. 
The parameters $a$, $b$, and $r_{c}$ are determined such that the potential and
its derivative are continuous at $r_{s}$ and $r_{c}$. 
The LJ/s model has essentially the same features as the LJ model, but since the potenial is of shorter range, the thermodynamic properties are different. The shorter range of the LJ/s also leads to significantly shorter simulation
times. 
Further details on the spline model and its thermodynamic properties can be found in refs. \cite{holian1983} and \cite{hafskjold2019}.

\begin{figure}[tbp]
\begin{center}
\includegraphics[trim=0 145 0 100, clip, width=1.0\linewidth]{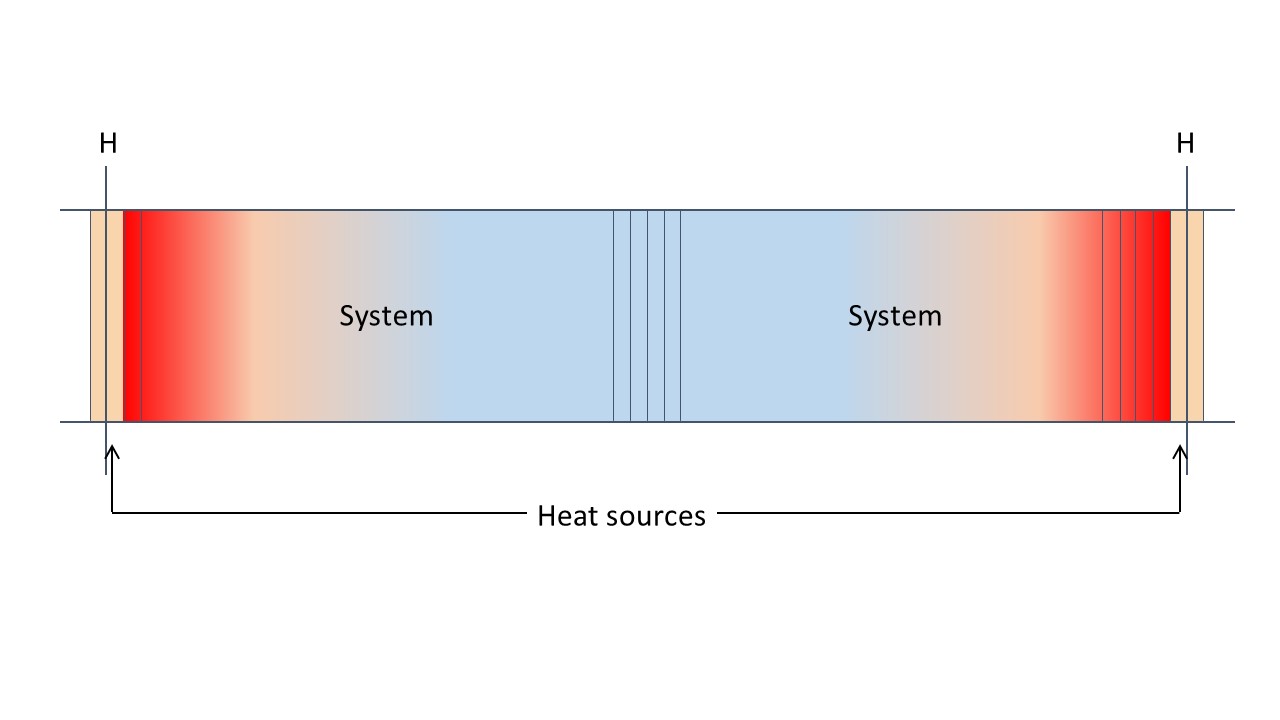}
\label{fig:sigma1}\end{center}
\caption{MD cell layout. The bright orange regions at the ends is where the energy was added during the heat pulse. The aspect ratio was different from that shown in the illustration.}
\label{fig:layout}
\end{figure}

The system layout is shown in Figure \ref{fig:layout}.
The simulations were made with a single component with $N=524,288$ particles in an elongated MD cell.
Periodic boundary conditions were used in all three directions.
The aspect ratio was set to $L_x/L_y=L_x/L_z=512$ in order to give the shock wave enough distance in $x$-direction to separate the wave front from the heat diffusion from the blast.
The number of layers was chosen so as to satisfy three criteria: (1) each layer should contain of the order 1,000 particles to ensure good signal-to-noise ratio for the properties computed in each layer, (2) the layer thickness, $\Delta x$, should be at least of the order one molecular mean free path, and (3) the resolution in $x$-direction should be good enough to see details of the density-, pressure-, and temperature profiles in the shock front, \textit{i.e.} $\Delta x$ should be at least 3-4 times smaller than the thickness of the shock front.
The overall number density was set to $n^*=N\sigma ^{3}/V=0.01$, where $V$ is the volume of the MD cell.\footnote{All numerical values throughout this paper are in dimensionless Lennard-Jones units and the corresponding symbol marked with an asterisk unless stated otherwise.} 
This low density allows us to use the virial expansion \cite{hafskjold2019} as an accurate equation of state in the analysis of the shock-wave data.
Condition (2) and (3) are counteractive in the sense that (2) favors a large $\Delta x$ whereas (3) favors a small $\Delta x$.
The thickness of the shock wave depends on its speed (the Mach number).
An estimate is $5-10$ times the molecular mean free path for Mach numbers $\approx 2$ and smaller for higher Mach numbers \cite{mott1951}.
An estimate of the mean free path based on elementary kinetic theory is $\lambda \approx \frac{V}{\sqrt{2} N \pi \sigma^2}$, which at the actual density amounts to approximately 20 in Lennard-Jones units.
The system was accordingly divided into 512 layers of equal thickness normal to the $x-$ direction, so that each layer contained on average of the order $1024$ particles.
The layers were used as control volume for computing local properties of the system.
With the density used in the simulations, this gives $\Delta x^* =29.5$, \textit{i.e.} approximately 50\% larger than the mean free path in the equilibrium gas ahead of the wave.

The blast was generated by thermostating one layer at each end of the MD cell to a temperature $T_\text{H}$ by simple velocity rescaling \cite{hafskjold1993}.
The other 510 layers were not thermostated.

The simulations included 20 parallel runs.
Each run was started from a configuration that was randomized with a Monte Carlo sequence of $m$ steps, $m=[1,2,...19,20]\times 10^{5}$ followed by equilibrium simulations at $T^\ast =k_{B}T/\varepsilon =1.0$.
This temperature is slightly above the critical temperature for this model ($T_\text{c}^\ast =0.885$, \cite{hafskjold2019}) and the gas has a Prandtl number $Pr \approx 1.4$. 
The number density and the mass density are numerically identical in reduced LJ units.
Each timestep was $\delta t^*=0.002$, with $t^*=\frac{t}{\sigma}\left ( \frac{\varepsilon}{m}\right )^{1/2}$. 
The density and temperature used in this work correspond to argon at approximately 120 K and 4 bar (assuming the usual Lennard-Jones parameters for $\varepsilon$ and $\sigma$, \textit{i.e.} $\varepsilon/k_B=124$ K and $\sigma = 3.418$ \AA).

Starting from the equilibrium state for each of the 20 equilibrated systems, energy was added as a pulse by setting the thermostats in the regions marked "H" in Fig. \ref{fig:layout} for 2,000 time steps.
Two cases were studied with $T_\text{H}^*=130$ and $2080$. 
This thermal blast and sudden increase in the local temperature and pressure at the ends of the MD cell generated pressure waves traveling in $x$-direction from the ends of the MD cell towards its center.
After the initial 2,000 time steps, the simulation was continued as a $NVE$-simulation with the same time step $\delta t^*$.
Since the 20 equilibrium configurations were slightly different from each other, the energy inputs were also different, leading to slightly different propagations of the individual waves.
This was particularly visible for the strong shock.
The symmetry of the system was used to pool data from the two halves of the cell.
The outcome of the 20 runs in each series was used for postprocessing of average properties and uncertainties.
\begin{figure}[tbp]
\centering
\includegraphics[scale=0.5]{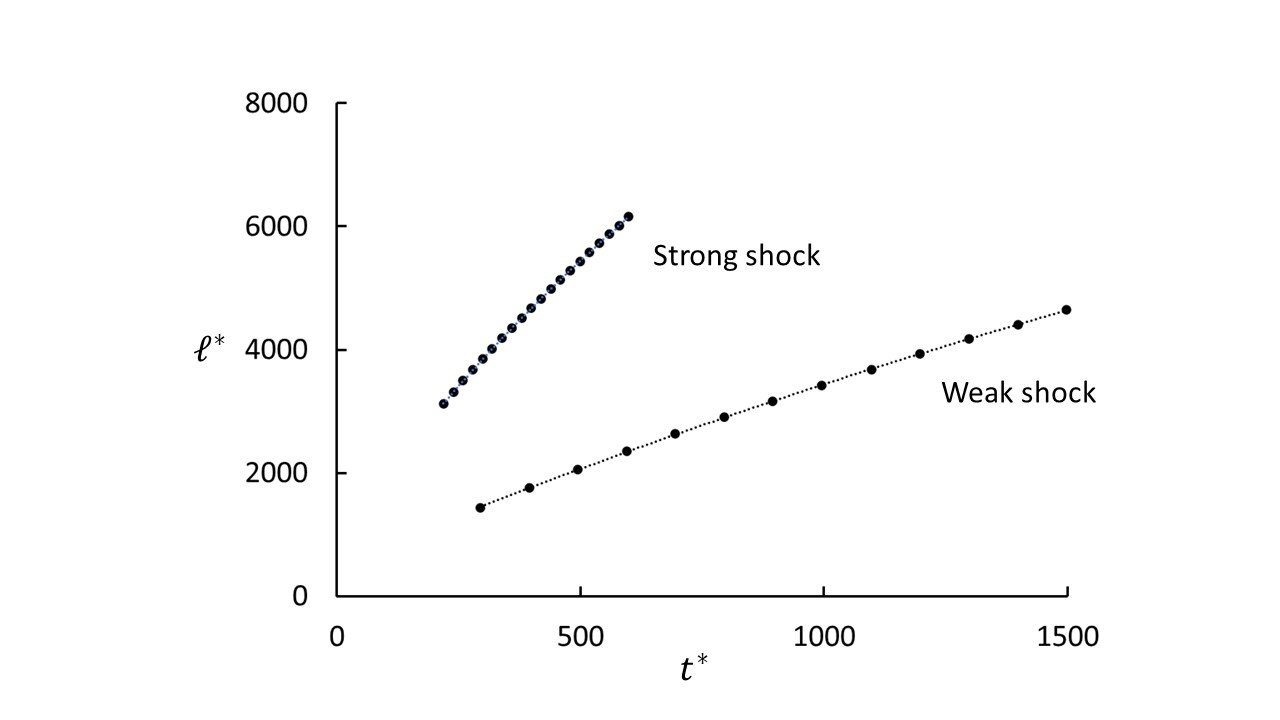}
\caption{Shock-front position as determined with the Gibbs' equal-area construction as function of time.
The dotted lines are third-order polynomial fits.}
\label{fig:waveposition}
\end{figure}
\begin{figure}[tbp]
\centering
\includegraphics[trim=0 30 0 50, clip, scale=0.5]{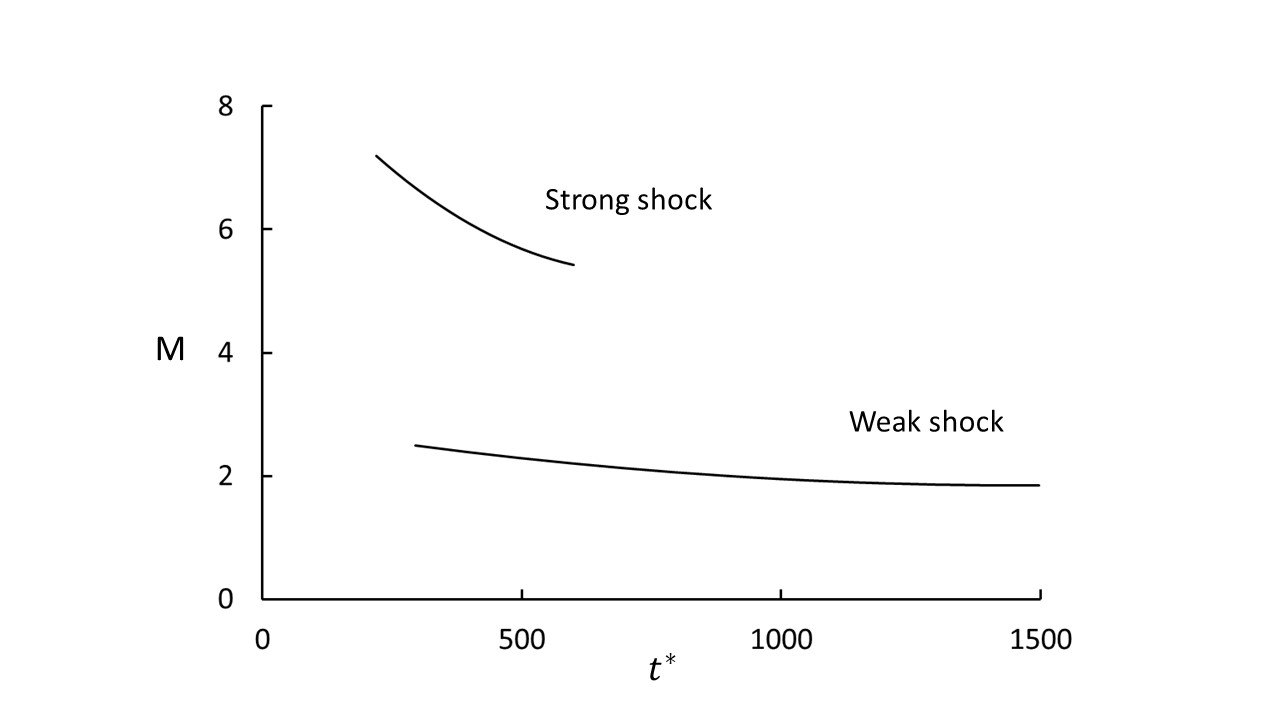}
\caption{Wave speed shown as Mach number as function of time for the two cases shown in \ref{fig:waveposition}.}
\label{fig:wavespeed}
\end{figure}

A typical density profile for the left half-cell is shown in Figure \ref{fig:gibbsequalarea} in reduced LJ units. 
The location of of the shock-wave front at a given time was determined from Eq.~\eqref{eqn:G.1} using the excess mass density $\rho ^\text{s}$ as illustrated in the insert in Figure \ref{fig:gibbsequalarea}. 
A linear function was fitted to the density profile $\rho (x)$ for $x < \ell$.
For $x > \ell$, the equilibrium density $\rho ^\ast =0.01$ was used in the extrapolation.
The condition $\rho^\text{s}(\ell )=0$, which determines the location $\ell $ of the shock front, was solved by the "solve" function in MS Excel and Simpson-integration of the NEMD data.

The position of the shock front was recorded as function of time and the wave speed was computed from Eq. \eqref{eqn:B.3}.
Figure \ref{fig:waveposition} shows the position as function of time for the strong shock generated in this work in comparison with the weak shock discussed in ref. \cite{hafskjold2020}.
Figure \ref{fig:wavespeed} shows that the speed decays with time, slowly for the weak shock and faster for the strong shock.
This indicates that energy is dissipated faster in the strong shock than in the weak shock.

The speed of sound in the gas ahead of the wave was determined from
\begin{equation}
v_\text{sound}^2=\frac{C_p}{C_v}\left ( \frac{\partial p}{\partial \rho} \right )_T
\end{equation}
by independent MD simulations of $C_p$, $C_V$, and $\left ( \partial{p}/\partial{\rho} \right )_T$ and found to be 1.298, which is essentially the ideal-gas value, 1.291, at $T^*=1.0$.

The NET-analysis of the entropy production requires information about the enthalpy, entropy, density, and kinetic energy in front of, and behind the shock wave.
In addition, we also need the transport properties mass flux, measurable heat flux, and the $x$-component of the viscous pressure tensor, which includes the shear and bulk viscosities of the gas. 
These properties were computed as time- and spacial averages of NEMD and equilibrium MD results using the expressions shown below.
The entropy was computed from the equation of state as explained in the supplementary material.

We have used the kinetic temperature as a measure of the temperature in our
analysis, $i.e.$ 
\begin{equation}
\textsf{T}=\frac{1}{3(N_{\text{CV}}-1)k_\text{B}}\sum_{i\in \text{CV}}m(\mathbf{v}_{i}-%
\mathbf{v})^{2}
\end{equation}%
where $k_\text{B}$ is Boltzmann's constant, $\mathbf{v}_{i}$ is the
3-dimensional velocity of particle $i$ (all the particles have the same mass, $m$, in this one-component case), and $\mathbf{v}$ is the streaming velocity (the velocity of the local center of mass).
The summation is done over all the $N_{\text{CV}}$ particles in the local control volume (CV), $i.e.$ each layer in the MD cell.
The local streaming velocity was determined as 
\begin{equation}
\mathbf{v}=\frac{1}{M_{\text{CV}}}\sum_{i\in \text{CV}}m\mathbf{v}_{i} = \frac{1}{N_{\text{CV}}}\sum_{i\in \text{CV}}\mathbf{v}_{i}
\label{eqn:streaming}
\end{equation}
where $M_{\text{CV}}=mN_{\text{CV}}$ is the total mass in CV. 
Because the transport is in $x$-direction only, the $y-$ and $z-$ components of $\mathbf{v}$\ are zero on average and the $x-$component is the local streaming velocity $v$, cf. Eqs. \eqref{eqn:cons1} - \eqref{eqn:cons3}. 

The kinetic temperature in a shock wave front has different values in $x$-, $y$-, and $z$-directions, and is therefore a tensorial quantity \cite{salomons1992}.
The temperature was first computed in the MD frame of reference, $\textsf{T}^{\text{MD}}=\left \{ \textsf{T}_{xx}^{\text{MD}}, \textsf{T}_{yy}^{\text{MD}}, \textsf{T}_{zz}^{\text{MD}} \right \}$ where
\begin{equation}
\textsf{T}_{qq}^{\text{MD}}=\frac{1}{(N_{\text{CV}}-1)k_\text{B}}\sum_{i\in \text{CV}}m v_{i,q}^2, \quad q=x, y, z
\label{eqn:Tqq}
\end{equation}
The conversion to the kinetic temperature was done in the postprocessing
using 
\begin{align}
\textsf{T}_{xx}& =\textsf{T}_{xx}^{\text{MD}}-\frac{M_{\text{CV}}}{N_{\text{CV}}k_\text{B}}
v^{2} = \textsf{T}_{xx}^{\text{MD}}-\frac{m }{k_\text{B}}v^{2}
\label{eqn:Txcorrected} \\
\textsf{T}_{yy}& =\textsf{T}_{yy}^{\text{MD}}  \label{eqn:Tycorrected} \\
\textsf{T}_{zz}& =\textsf{T}_{zz}^{\text{MD}}  \label{eqn:Tzcorrected}
\end{align}%
All these quantities are local in the CV.
The reported data for these quantities are space- and time averages of these quantities.

The shock wave creates a sharp density gradient in the fluid.
The pressure was therefore calculated using the coarse-grained version of the virial equation \cite{walton1983,ikeshoji2003} with the Irving-Kirkwood contour $C_{ij}$, the straight line between $i$ and $j$ \cite{irving1950}.
We summarize the method here for a plane surface normal to the $x$-direction.
Consider a pair of particles $ij$.
One of them or both may be either inside or outside CV.
The configurational contribution to the $qq$-component of the pressure in CV from that pair is
\begin{equation}
\textsf{P}_{\text{conf},qq}=\frac{1}{2} \sum_{i=1}^N \sum_{\substack{ j=1 \\ j \ne i}}^N \textsf{P}_{ij,qq}
\label{eqn:pconf}
\end{equation}%
where
\begin{equation}
\textsf{P}_{ij,qq}=-\frac{1}{V} \int_{\text{CV}} \left [ \int_{C_{ij}} f_{ij,q} \delta (\mathbf{R}-\mathbf{l})dl_q\right ]d\mathbf{R}
\label{eqn:IKintegral}
\end{equation}%
where $\mathbf{R}$ is some point in space and $\mathbf{l}$ is a point on the contour $C_{ij}$.
In the present context, Eq. \eqref{eqn:IKintegral} reduces to
\begin{equation}
\textsf{P}_{ij,qq}=-\frac{f_{ij,q} r_{ij,q}}{V r_{ij,x}} H(x_i,x_j)
\label{eqn:IKintegral2}
\end{equation}
where $H(x_i,x_j)$ is a book-keeping function that defines how much of the contour $C_{ij}$ that is inside the control volume.
Further details of the algorithm were described by Ikeshoji \textit{et al.} \cite{ikeshoji2003}.
The kinetic contribution to the pressure was computed as
\begin{equation}
\textsf{P}_{\text{kin},qq}= \frac{N k_\text{B} \textsf{T}_{qq}}{V}
\end{equation}

In general, the fluxes depend on how we choose the frame of reference.
In this context, there are three obvious choices, the MD cell coordinate system, the barycentric coordinate system, and the shockwave co-moving coordinate system. 
In MD simulations, fluxes are most conveniently computed in the MD cell (stationary) frame of reference.
The mass flux $j$ defined by Eq. \eqref{eqn:j} refers to the co-moving coordinate system and the total heat flux in Eq. \eqref{eqn:C.5} refers to the barycentric frame of reference.
Conversion between different frames of reference was done in postprocessing as described in the following.

The local streaming velocity is given by Eq. \eqref{eqn:streaming}.
The total heat flux in $x$-direction in the barycentric frame of reference was given by Evans and Morriss \cite{evans1990}: 
\begin{equation}
\label{eqn:heatflux}
J_{q,x}=\frac{1}{V_{\text{CV}}}\sum_{i\in \text{CV}}\left[ \left( 
\frac{1}{2}m ( \mathbf{v}_{i}-\mathbf{v} )^{2}+\phi
_{i}\right) \left( v_{i,x}-v\right) -\frac{1}{2}%
\sum_{\substack{ j=1 \\ j \ne i}}^{N}\left[ \left( \mathbf{v}_{i}-\mathbf{v}\right)
\boldsymbol{\cdot} \mathbf{f}_{ij}\right] x_{ij}\right] 
\end{equation}
where $\phi _{i}$ is the potential energy of particle $i$ in the field of
all the other particles within range (including those outside the $\text{CV}$%
), $\mathbf{f}_{ij}$ is the force acting on $i$ due to $j$, and $x_{ij}=x_j - x_i$ is the distance from $i$ to $j$ in $x$-direction.
The total heat flux in the barycentric frame of reference is equal to the measurable heat flux in the one-component system considered here. 

The corresponding energy flux in the MD cell frame of reference is found by setting $v=0$ and $\mathbf{v}=\{0,0,0\}$ in Eq. \eqref{eqn:heatflux}:
\begin{equation}
\label{eqn:MDheatflux}
J_{q,x}^{\text{MD}}=\frac{1}{V_{\text{CV}}}\sum_{i\in \text{CV}}\left[ \left( \frac{1}{2}m \mathbf{v}_{i}^2+\phi_{i}\right) v_{i,x} -\frac{1}{2} \sum_{\substack{ j=1 \\ j \ne i}}^{N}\left( \mathbf{v}_{i} \boldsymbol{\cdot} \mathbf{f}_{ij}\right) x_{ij}\right] 
\end{equation}
Eq. \eqref{eqn:MDheatflux} introduced into Eq. \eqref{eqn:heatflux} allows a separation of the heat flux into $J_{q,x}^{\text{MD}}$ and the rest:
\begin{equation}
J_{q,x}=J_{q,x}^{\text{MD}}-J_{q,x}^{\text{flow}}
\end{equation}
where
\begin{align}
\label{eqn:flowheatflux}
J_{q,x}^{\text{flow}}=&\frac{v}{V_{\text{CV}}}\left[ \sum_{i\in \text{CV}}\left( \frac{1}{2}m \mathbf{v}_{i}^2+\phi_{i}\right) + \sum_{i\in \text{CV}}m_i v_{i,x}^2 -\frac{1}{2}\sum_{i\in \text{CV}} \sum_{\substack{ j=1 \\ j \ne i}}^{N}f_{ij,x}x_{ij} \right ] - \frac{v^3}{V_{\text{CV}}}\sum_{i\in \text{CV}} m_i \nonumber \\
=&v \left (\rho u+\textsf{P}_{xx} - \rho v^2 \right )
\end{align}

\section{Results and discussion}

\label{results}

In this Section, we first discuss our findings for the kinetic properties, \textit{viz.} the kinetic temperature and the velocity distributions.
We show that the temperature is non-isotropic, in agreement with previous results \cite{holian2010}.
In Section \ref{pressureprofiles}, we include the potential energy and the configurational contribution to the pressure and show that the non-equilibrium properties deviate from the equilibrium values in the microscopic description of the shock front.
Section \ref{entropybalance} is devoted to the entropy production computed by BBM and SBM.
The GEM is a major contribution in this work and will be discussed in detail in Section \ref{Gibbs}, including an analysis of the energy conversions in the shock front.
Finally, the four methods are compared in Section \ref{comparison}, where we discuss the validity of our calculation of the entropy production.

\subsection{Temperature and velocity profiles}
\label{velocityprofiles}

\begin{figure}[tbp]
\includegraphics[trim=0 30 0 0, clip, width=0.9 \columnwidth]{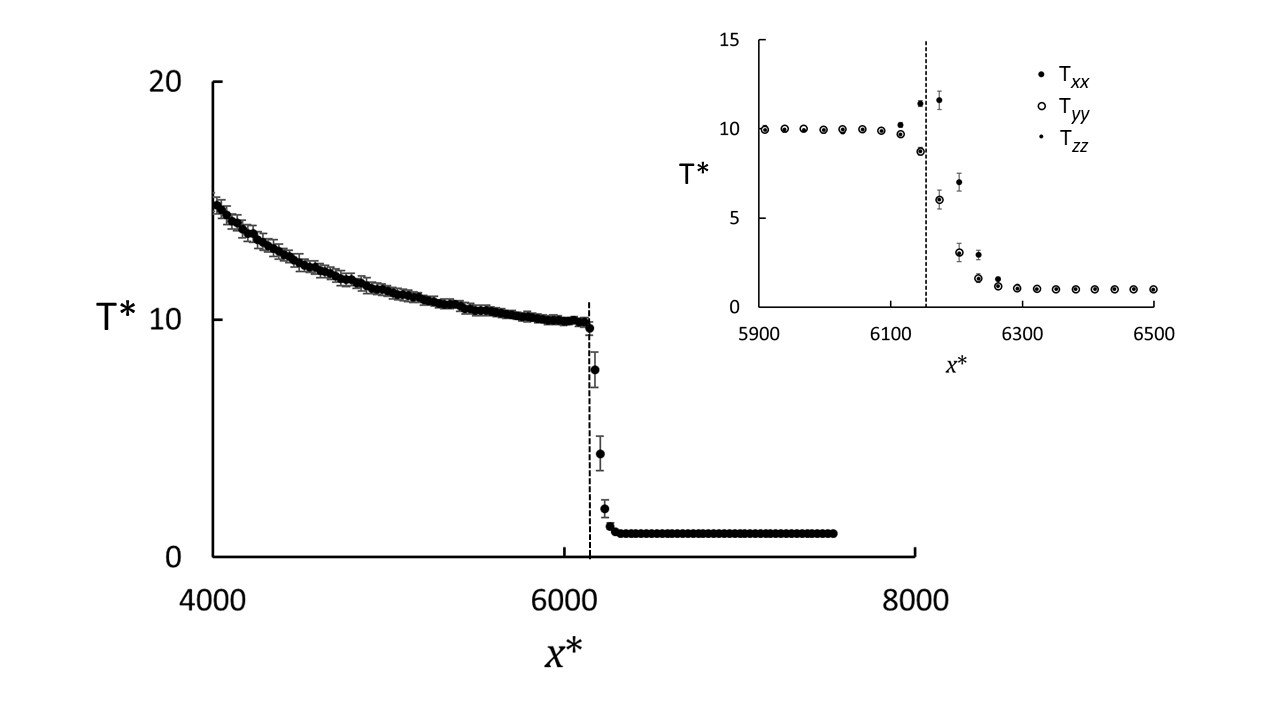}
\caption{Profile of $\textsf{T}=\frac{1}{3}\left ( \textsf{T}_{xx}+\textsf{T}_{yy}+\textsf{T}_{zz} \right )$ for the strong shock at time $t^*=600$. 
The insert shows that the normal and tangential components of the temperature tensor are different in the shock front, but equal immediately behind the shock. The uncertainties are three standard errors. The vertical dashed line shows the position of the equimolar surface.}
\label{fig:temperature}
\end{figure}

Shock waves are non-equilibrium and non-isotropic structures.
For instance, the kinetic temperature in the shock front is non-isotropic as shown in previous simulations \cite{hoover2015,holian1993,holian2010,holian1983,hoover2014,hoover2016}.
The insert of Fig. \ref{fig:temperature} agrees with these earlier simulations;  the kinetic temperature is highly non-isotropic in the front of the strong shock, which indicates lack of local\footnote{We remind the reader that the term "local" refers to a control volume in the simulation cell, \textit{i. e.} a layer of thickness $\Delta x$.} equilibrium in the system.
A peak in $\textsf{T}_{xx}$ is known to occur for strong shocks \cite{holian1993,holian2010}.
In this work, we have used $T=\frac{1}{3}\text{Tr}(\textsf{T})$ throughout.

\begin{figure}[tbp]
\includegraphics[width=0.9 \columnwidth]{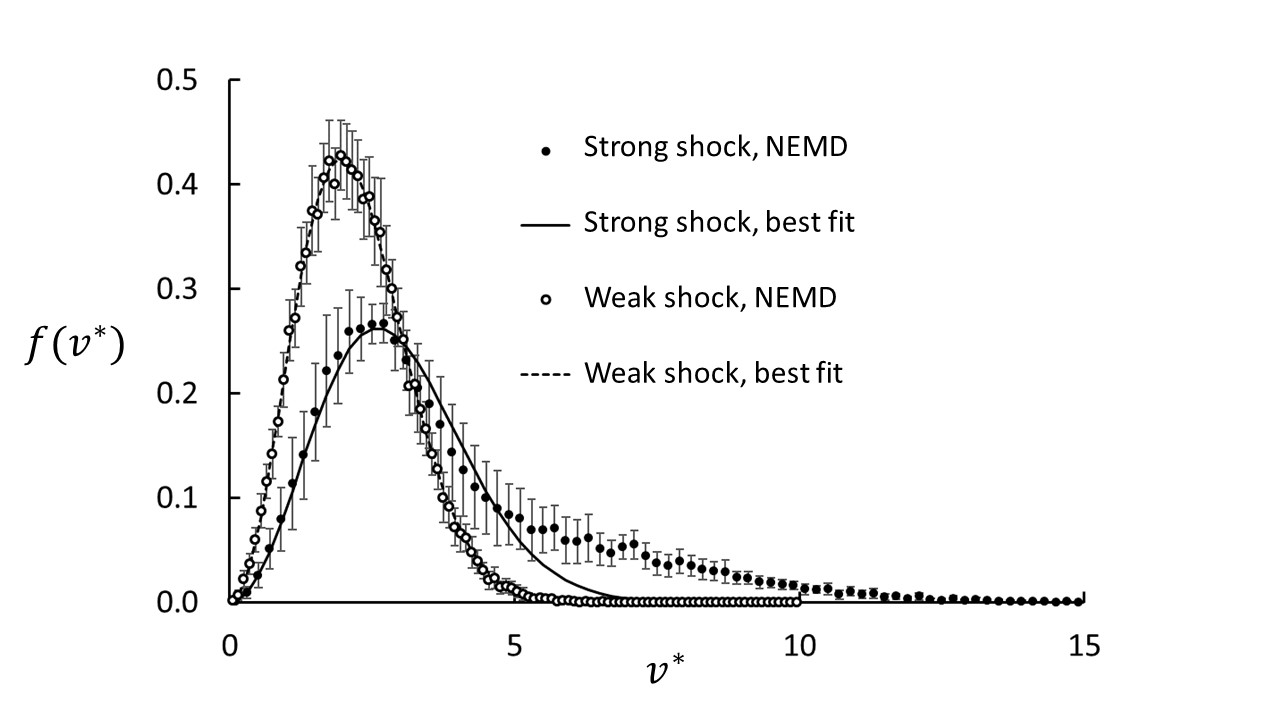}
\caption{Particle speed distributions for the weak and strong shocks.
The data for the weak shock were recorded at $x^*=3434$ and $t^*=1,000$, and for the strong shock at $x^*=6204$ and $t^*=600$.
The weak shock shows a perfect Maxwell-Boltzmann distribution, whereas the strong shock does not. The uncertainties are three standard errors.}
\label{fig:maxwell}
\end{figure}

We showed in a recent paper that the speed distribution in a weak shock front (Mach number 2.1) was a perfect Maxwell-Boltzmann distribution and concluded that this was consistent with a state of local equilibrium \cite{hafskjold2020}.
A comparison of the distribution functions for the weak and strong shocks is illustrated in Fig.~\ref{fig:maxwell}, based on the speed of $N_\text{CV} \sim 30,000$ particles (total from 20 runs) that were in a control volume of thickness $\Delta x^*$, centered at positions $x^*=3434$ and $6204$ for the weak and strong shock, respectively, and at the end of each simulation run.\footnote{The local streaming velocity was subtracted from $v_x$ in this analysis.}
The mean free path is $\lambda^* \approx 2\Delta x^*/3$ ahead of the wave and $\lambda^* \approx \Delta x^*/3$ behind the wave.
The fitted Maxwell-Boltzmann distribution gave a temperature $T^* = 1.92 \pm 0.01$ for the weak shock, in fair agreement with the local kinetic temperature $T^* = 1.79 \pm 0.01$.
The corresponding numbers for the strong shock are $T^* = 3.4 \pm 0.2$ from the fitted distribution, in poor agreement with the local kinetic temperature $T^*=5.1 \pm 0.7$ (uncertainties given as three standard errors of the mean).
%

To compute the surface excess entropy production with the GEM boils down to using Eqs. \eqref{eqn:N}-\eqref{eqn:Nb}.
It is worth noting that apart from the variable $T^\text{s}$ in these equations, the values of all the other properties are extrapolated values from the regions ahead of, and behind the shock front.
The system ahead of the shock is in local (and global) equilibrium.
The system immediately behind the shock is also in local equilibrium as shown by the temperature profiles.
For the purpose of this work, we conclude that, despite the fact that the system in each control volume in the shock front is not in local equilibrium for the strong shock, the thermodynamic properties used in the GEM (adjacent to the shock front) are in local equilibrium.

\subsection{Energy and pressure profiles}
\label{pressureprofiles}

In this section, we consider the configurational contributions to local properties, in particular the internal energy and pressure.
The system in question in this work is a moderately dense gas, so the configurational contributions are likely to be small.
We will nevertheless assess to what extent the non-equilibrium configurational properties deviate from the equilibrium values with focus on the mechanical properties internal energy and pressure, which are easily obtained in NEMD.
Irrespective of the main theme of this work, \textit{viz.} the entropy production, this assessment provides interesting insight in the shock wave by itself.
\begin{figure}[tbp]
\includegraphics[trim=0 40 0 30, clip, width=1.0\columnwidth]{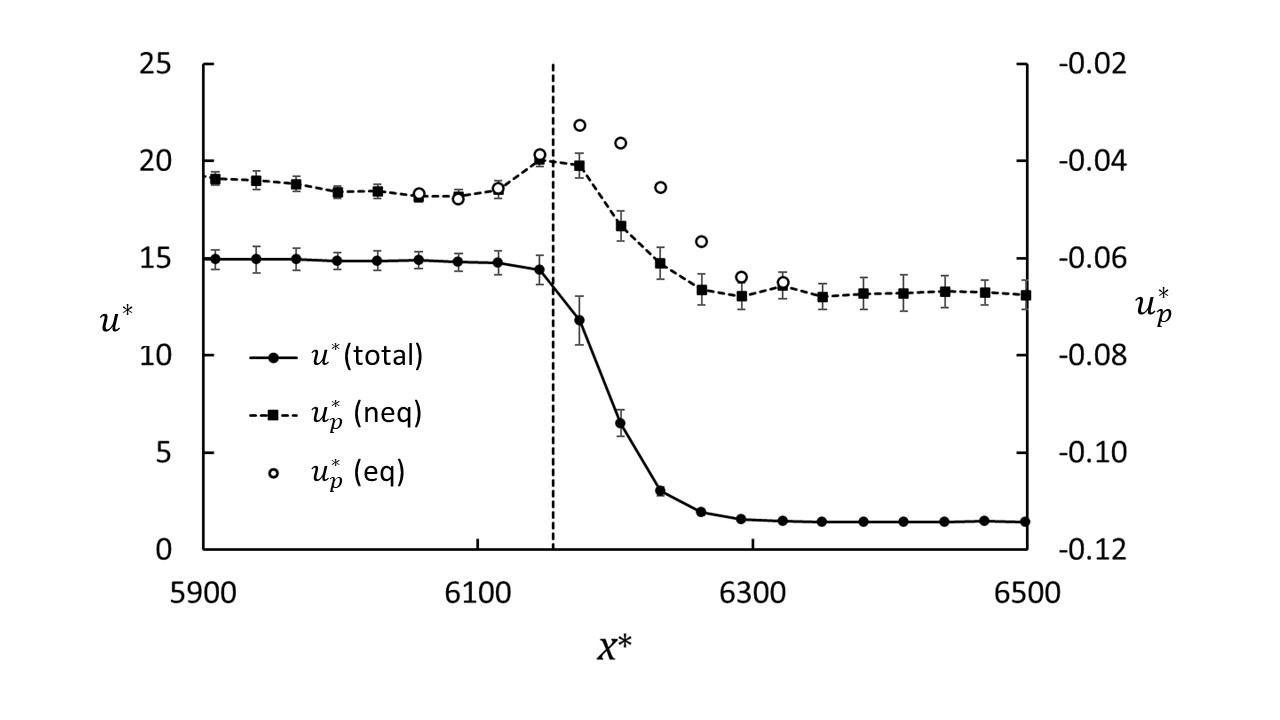}
\caption{Specific internal energy as function of $x$ in the front region of the strong shock at $t^*=600$.
The dots show the total internal energy determined by NEMD, $u^*$(total), and the squares show the configurational (potential) part of it, $u_p^*$.
The black and white squares show data from the non-equilibrium and equilibrium simulations, respectively.
Note that the configurational contributions (referring to the right axis) are so small that they do not visibly separate the kinetic contributions to the internal energy from the total in $u^*$.
The error bars are three standard errors, the errors for the equilibrium results are smaller than the symbol size. 
The vertical dashed line shows the position of the Gibbs equimolar surface.
}
\label{fig:energyprofile}
\end{figure}

The specific internal energy, $u$, for the strong shock is shown in Figure \ref{fig:energyprofile}.
The internal energy is completely dominated by the kinetic part, the potential (configurational) energy accounts for at most $\approx 0.3 \%$ of the total internal energy.\footnote{The potential energy contribution will clearly be larger at higher densities, such as in a liquid.}
The potential energy is less negative at the downstream side of the surface because the particles are on average closer together there.
The potential energy does show a difference between the non-equilibrium energy and the energy determined by equilibrium simulations in the range $6,200 \leq x^* \leq 6,300$.
The difference is $\approx 40 \%$ of the potential energy or $\approx 0.1 \%$ of the total internal energy.
The equilibrium data were generated at the local non-equilibrium density and temperature $T=\frac{1}{3}\text{Tr}(\textsf{T})$ with the temperature components given by Eqs. \eqref{eqn:Txcorrected} - \eqref{eqn:Tzcorrected}.
It is interesting to note that this difference occurs at the upstream, low-density side of the equimolar surface.
Based on this result, we conclude that the internal energy density is very accurately given by the equilibrium values.

\begin{figure*}[!ht]
\centering
  \subfloat[]{\includegraphics[trim=0 30 0 0, clip, width=0.9\columnwidth]{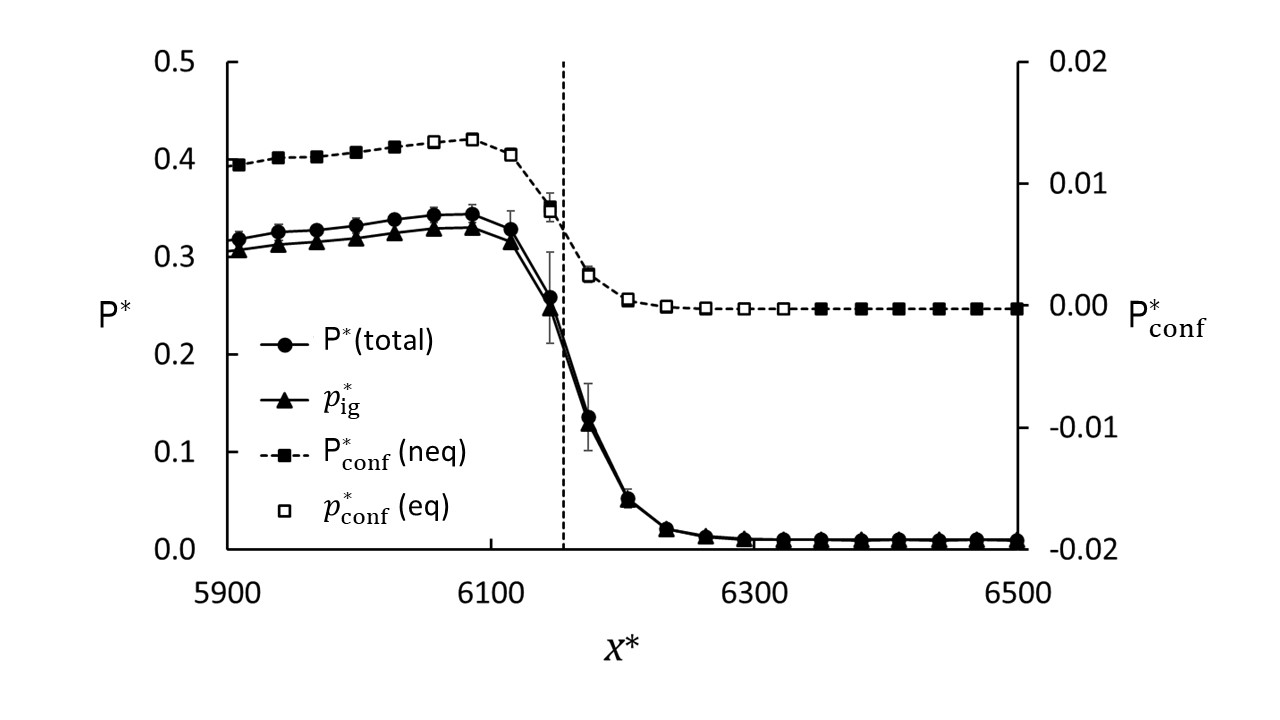}
  \label{fig:pressure1}}
\\
  \subfloat[]{
  \includegraphics[trim=0 30 0 50, clip, width=0.9\columnwidth]{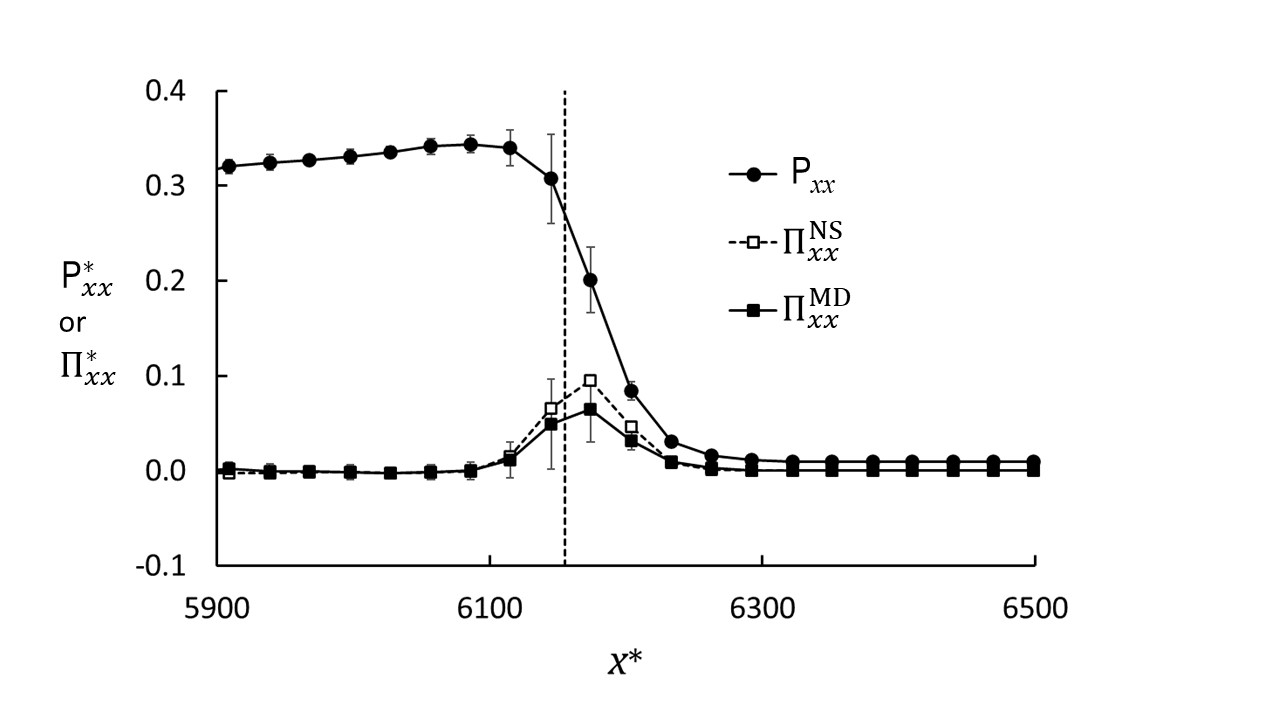}
  \label{fig:pressure2}}\\
\caption{Contributions to the pressure for $5900 \leq x^* \leq 6500$ at $t^*=600$ for the strong shock.
The vertical dashed line shows the position of the Gibbs equimolar surface at $x^*=6154$.
\textbf{(a)} The black dots and triangles represent the total non-equilibrium pressure, $\frac{1}{3}\text{Tr}(\textsf{P})$, and its kinetic contributions, $p_\text{ig}$, respectively.
The squares are the corresponding equilibrium pressures $p$ and $p_\text{conf}$.
Note that the kinetic contributions are equal in the two cases.
\textbf{(b)} The $x$-component of the pressure tensor ($\textsf{P}_{xx}$) and the viscous pressure computed from the NEMD data ($\Pi_{xx}^\text{MD}$) and Navier-Stokes ($\Pi_{xx}^\text{NS})$.
}
\label{fig:tempz}
\end{figure*}

Figure \ref{fig:pressure1} shows the total pressure $\frac{1}{3}\text{Tr}(\textsf{P})$ as function of $x$ in the shock front region for the strong shock at $t^*=600$.
The non-equilibrium pressure is, assuming the Navier-Stokes relations, $\frac{1}{3}\text{Tr}(\textsf{P}) + \eta_\text{B} \frac{\partial v_x}{\partial x} \approx \frac{1}{3}\text{Tr}(\textsf{P})$.
The bulk viscosity $\eta_\text{B}$ is small for a monatomic dilute gas.
We have estimated, based on data from Hoheisel \textit{et al.} for the Lennard-Jones fluid \cite{meier2005}, that $\eta_\text{B}^* < 10^{-3}$ for our Lennard-Jones spline system in the actual states, which makes the contribution from the bulk viscosity to the pressure at least four orders of magnitude smaller than $\frac{1}{3}\text{Tr}(\textsf{P})$.
We have therefore assumed that $\eta_\text{B}=0$ in this work.
The total pressure may be split into a kinetic (ideal-gas) contribution, $\frac{1}{3}\text{Tr}(\textsf{P}_\text{kin})$ and a configurational term, $\frac{1}{3}\text{Tr}(\textsf{P}_\text{conf})$.
For the kinetic term, we have used (in Lennard-Jones units) $\frac{1}{3}\text{Tr}(\textsf{P}_\text{kin}^*) = \rho^*\frac{1}{3}\text{Tr}(\textsf{T}^*)=\rho^*T^*=p_\text{ig}^*$.
The configurational term was computed according to Eqs. \eqref{eqn:pconf}-\eqref{eqn:IKintegral2}.
The pressure is almost zero ahead of the shock and increases monotonically through the front.
The ideal-gas pressure accounts for 98 \% and 96 \% of the total pressure immediately upstream and downstream, respectively, of the shock.
Figure \ref{fig:pressure1} also shows a comparison between the non-equilibrium and equilibrium configurational pressures.
The configurational pressure is slightly negative ahead of the shock wave and positive behind the wave where the gas is more compressed.
Unlike the configurational energy, there is virtually no difference between the equilibrium and non-equilibrium pressures.
This is consistent with the Navier-Stokes relations, Eqs. \eqref{eqn:const1} and \eqref{eqn:const2}, which imply that $\frac{1}{3}\text{Tr}(\textsf{P}) = p - \eta_\text{B} \frac{\partial v_x}{\partial x} \approx p$.

Eqs. \eqref{eqn:const1} and \eqref{eqn:const2} may be combined to $\Pi_{xx}^\text{MD} \approx p$ for the viscous pressure, where superscript "MD" means "as determined from the NEMD simulations".
The viscous pressure may also be determined as $\Pi_{xx}^\text{NS}=-\left ( \frac{4}{3}\eta_\text{S} + \eta_\text{B} \right ) \frac{\partial v_x}{\partial x} \approx - \frac{4}{3}\eta_\text{S} \frac{\partial v_x}{\partial x}$ where we have used superscript "NS" to distinguish it from the NEMD results.
The shear viscosity $\eta_\text{S}$ was determined by independent non-equilibrium MD simulations with LAMMPS \cite{plimpton1995} and $\frac{\partial v_x}{\partial x}$ was taken from the velocity profile in the shockwave.
Fig. \ref{fig:pressure2} shows a comparison between $\Pi_{xx}^\text{MD}$ and $\Pi_{xx}^\text{NS}$.
The main observation is that $\Pi_{xx}$ contributes only in the shock front where $\frac{\partial v_x}{\partial x}$ is significant.
Here, $\Pi_{xx}$ accounts for up to 40\% of $\textsf{P}_{xx}$.
The dominant viscous contribution is on the low-density side of the equimolar surface.
The agreement between $\Pi_{xx}^\text{MD}$ and $\Pi_{xx}^\text{NS}$ is within the uncertainties, and again consistent with the Navier-Stokes equations.
This indicates that the Navier-Stokes equation gives a correct description of the pressure profiles.

\subsection{The entropy production}
\label{entropybalance}

Based on the analyses in sections \ref{velocityprofiles} and \ref{pressureprofiles}, we found that the assumption of local equilibrium is good as measured by the internal energy and pressure.
Lacking values for the non-equilibrium entropy, we shall in the following assume that also the entropy can be estimated by the equilibrium values.
We shall now use these results to determine the surface excess entropy production with the four methods; the BBM based on Eq. \eqref{eqn:B.1}, the LIT based on Eq. \eqref{eqn:B.1a}, the SBM based on Eq. \eqref{eqn:S.1}, and the GEM based on Eqs. \eqref{eqn:N} - \eqref{eqn:Nb}.

\subsubsection{The bulk balance method (BBM) and linear irreversible thermodynamics (LIT)}

\begin{figure*}[!ht]
\centering
  \subfloat[]{\includegraphics[trim=0 30 0 0, clip, width=0.9\columnwidth]{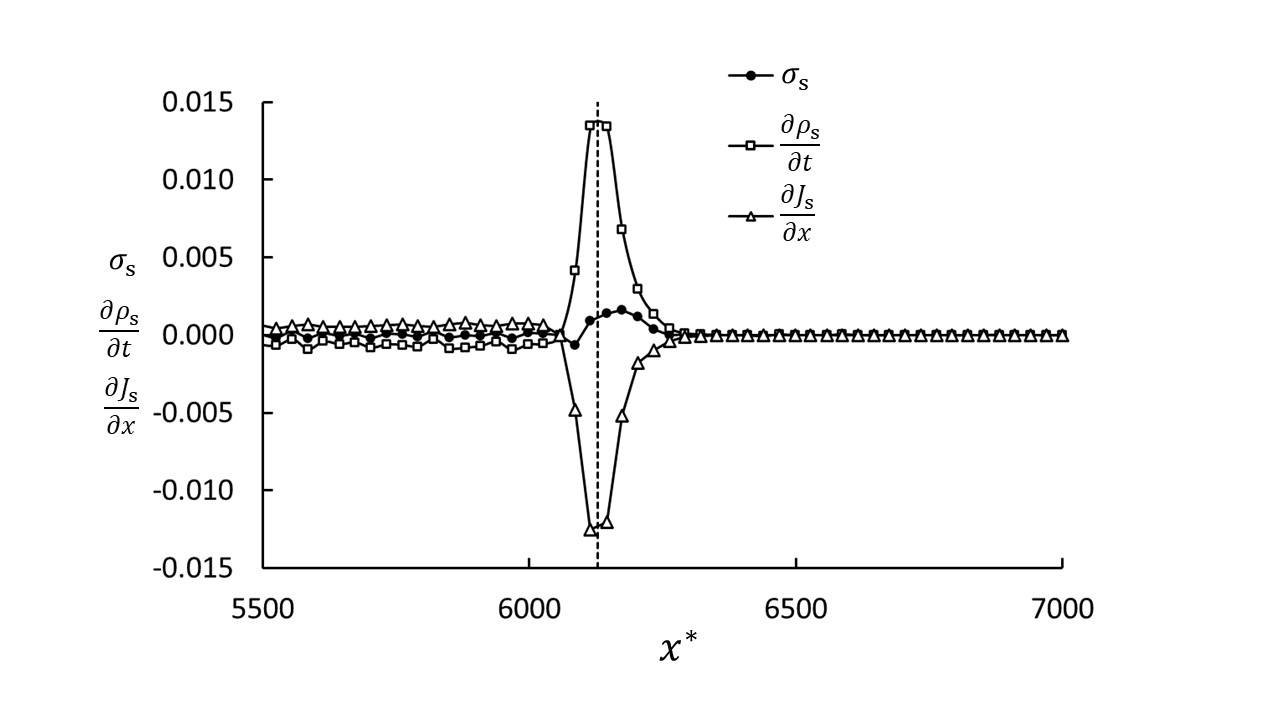}
  \label{fig:BBM2}}
\\
  \subfloat[]{
  \includegraphics[trim=20 30 0 50, clip, width=0.85\columnwidth]{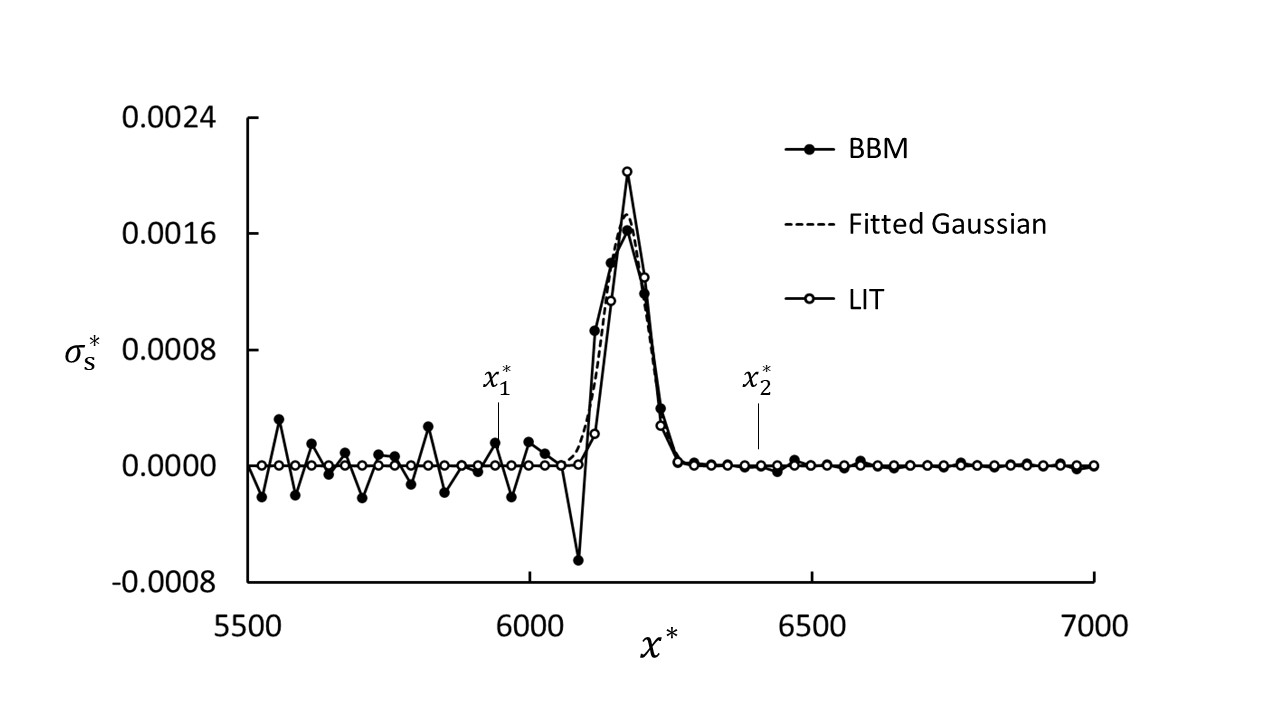}
  \label{fig:BBM1}}\\
\caption{Data used in the BBM for the strong shock at $t^*=600$. Uncertainties have not been determined, but are probably similar to the scatter on the left of the wave.
(a) The functions $\frac{\partial}{\partial t} \rho_\text{s}(x,t)$ and $\frac{\partial }{\partial x} J_\text{s}(x,t)$ and their sum $\sigma_\text{s}(x,t)$. The vertical dashed line is the position of the Gibbs equimolar surface.
(b) The integrand $\sigma_\text{s}(x,t)$ and the fitted Gaussian. We also show results from the LIT method used by Velasco and Uribe \cite{velasco2019}, \textit{i.e.} the Gibbs equation used locally for each control volume in the simulation (see text).}
\label{fig:BBM}
\end{figure*}

Integrating Eq. \eqref{eqn:B.1} over the thickness of the wave, we get the total entropy production in the wave, which is also the surface excess entropy production:
\begin{equation}
\sigma_\text{s}(t)= \int_{x_1}^{x_2} \sigma_\text{s}(x,t) dx 
\label{eqn:eq7}
\end{equation}
In this method, we consider the properties of the wave as continuously changing over the wave front like in the bulk.
The terms $\frac{\partial}{\partial t} \rho_\text{s}(x,t)$ and $\frac{\partial }{\partial x} J_\text{s}(x,t)$ in Eq. \eqref{eqn:B.1} were determined by five-point numerical differentiation with the results shown in Figure \ref{fig:BBM2}.
The two contributions are opposite in sign with a relatively small sum.
The integrand $\sigma_\text{s}(x,t)$ and the integration limits $x_1$ and $x_2$ are shown in Figure \ref{fig:BBM1}.
A Gaussian function was fitted to $\sigma_\text{s}(x,t)$ to smooth the NEMD data and the fit was integrated analytically.
The graph shows that the values of the integration limits were not critical, the entropy production occurs only in the shock front.
The negative dip in $\sigma_\text{s}(x,t)$ at the left side of the peak in Figure \ref{fig:BBM1} is due to a slight mismatch in the peaks of $\frac{\partial}{\partial t} \rho_\text{s}(x,t)$ and $\frac{\partial }{\partial x} J_\text{s}(x,t)$.
We believe this is not significant and an artifact of the five-point numerical differentiation methods we have used.
The results from the BBM show that the entropy production is negligible a few mean free paths away from the shock front.
The surface represents the dominant entropy production.
This procedure was repeated for a series of times between $t^*=200$ and $t^*=600$. 
The surface excess entropy production is shown as function of time in Figure \ref{fig:finalresult} and compared with data from the other three methods used.

We showed in Sections \ref{velocityprofiles} and \ref{pressureprofiles} that although the system is not in local equilibrium in the shock front region, it is close to being so.
This is the basis for using the Gibbs equation in the normal way for bulk fluids \cite{deGroot1962}, like Velasco and Uribe did \cite{velasco2019}.
The key result for the entropy production in this method is Eq. \eqref{eqn:B.1a}, which integrated over the shock thickness (Eq. \eqref{eqn:eq7}) gives $\sigma_\text{s}^\text{s}$.
The $\sigma_\text{s}(x)$ determined in this way is shown in Figure \ref{fig:BBM1} marked "LIT".
The agreement with the BBM is very good, indicating that the local equilibrium assumption is good even for the strong shock.
Whereas $\sigma_\text{s}(x)$ determined from Eq. \eqref{eqn:B.1} is a small difference between large numbers (\textit{cf.} Figure \ref{fig:BBM2}), when determined from Eq. \eqref{eqn:B.1a}, it is a sum of small numbers and therefore less noisy, especially downstream of the shock front.
All the quantities on the RHS of Eq. \eqref{eqn:B.1a} were determined directly from the NEMD results without any assumptions for the transport coefficients.
Integrating Eq. \eqref{eqn:B.1a} over the entire system showed that some 97 \% of the total entropy production occurred in the shock front.


\subsubsection{The surface balance method (SBM)}

\begin{figure}[!ht]
\begin{center}
\includegraphics[trim=30 30 0 0, clip, width=1.0\columnwidth]{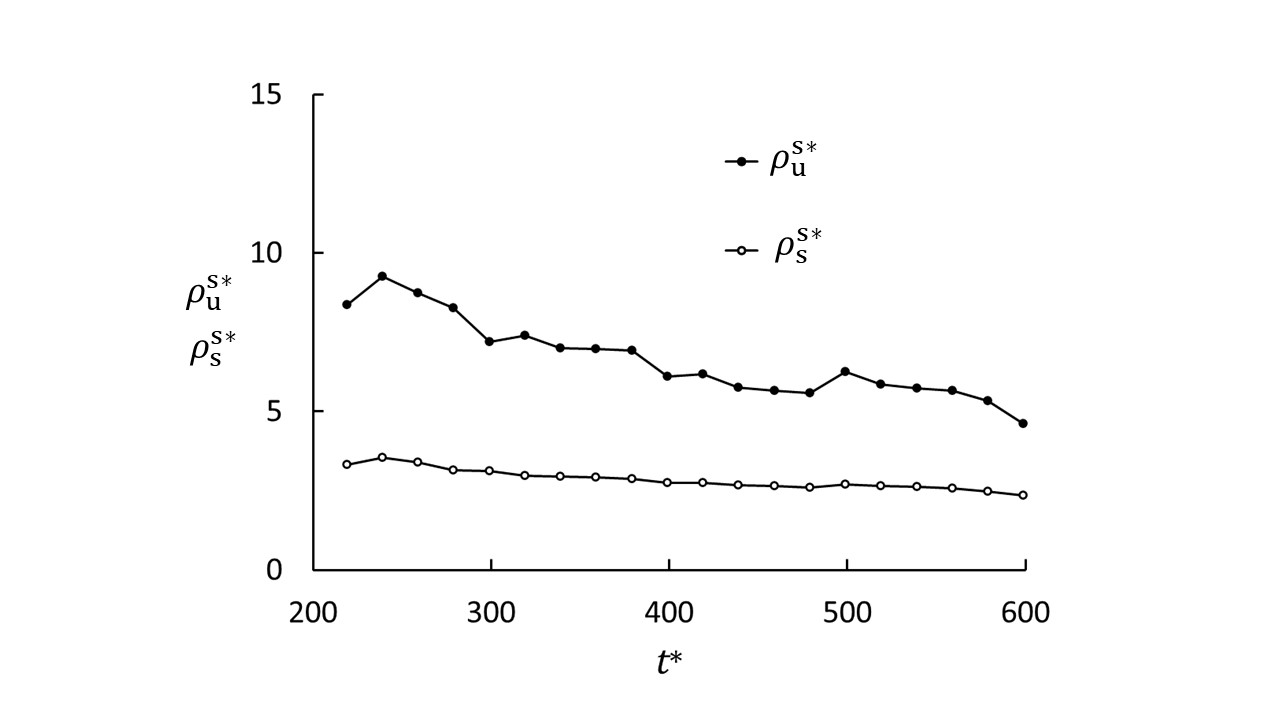}
\end{center}
\caption{Surface excess entropy density and internal energy density as function of time for the strong shock.}
\label{fig:rhosrhou}
\end{figure}
In the derivation of Eq. \eqref{eqn:S.1}, we considered the wave front as a surface, but without employing the Gibbs equation.
The term $\rho_\text{s}^\text{s}(t)$ was determined from Eq. \eqref{eqn:G.1} with the result shown as function of time in Figure \ref{fig:rhosrhou}.
As input to Eq. \eqref{eqn:G.1}, we used the entropy density given by the equation of state, using the local density and temperature as input.
The equation of state we used was based on the virial expansion and is given in the supplementary information.
The time derivative was determined from a linear fit to $\rho_\text{s}^\text{s}(t)$.
The surface excess entropy density varies little with time and contributes less than 2 \% to $\sigma_\text{s}^\text{s}$.

The term $[J_\text{s}]_-$ depends on the extrapolated values from properties outside the shock front, which we have shown in Sections \ref{velocityprofiles} and  \ref{pressureprofiles} to be well represented by equilibrium values.
Because of this, we consider Eq. \eqref{eqn:S.1} to give a reliable estimate for the surface excess entropy production.
Results are compared with the other three methods in Figure \ref{fig:finalresult}.

\subsubsection{The Gibbs excess method (GEM)}
\label{Gibbs}

\begin{figure*}[!ht]
\centering
\includegraphics[width=1.0\columnwidth]{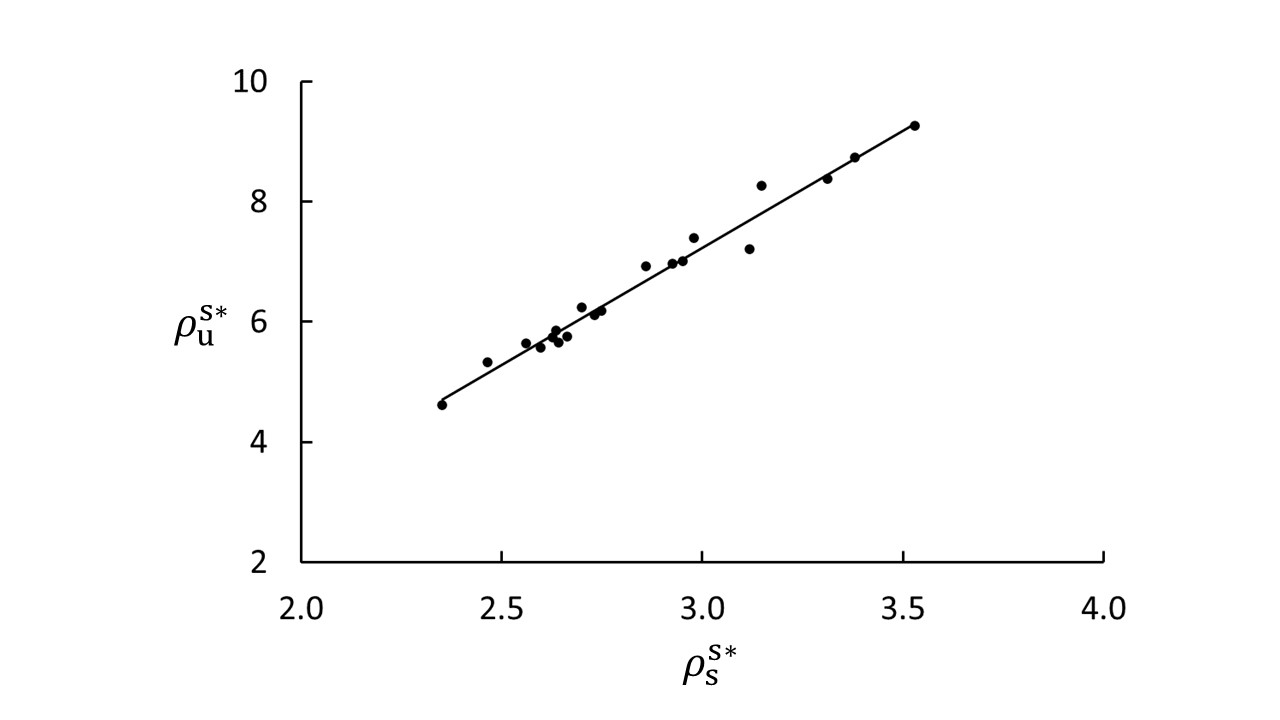}
\caption{ Surface excess internal energy density vs. surface excess entropy density. The slope was determined to $T^{\text{s}*}=3.9$.}
\label{fig:Ts}
\end{figure*}

In the GEM, the surface excess entropy production is determined from Eqs. \eqref{eqn:N} - \eqref{eqn:Nb}.
All quantities in these equations, except the surface temperature $T^\text{s}$, are determined by extrapolating properties from the bulk phases adjacent to the wave front.
We established in Sections \ref{velocityprofiles} and \ref{pressureprofiles} that these properties are given by their equilibrium values in the present case.
The surface temperature is, however, given by Eq. \eqref{eqn:S.3}.
This equation involves the excess properties $\rho_\text{u}^{\text{s}}$ and $\rho_\text{s}^{\text{s}}$ determined from the entire profiles, including the non-equilibrium properties in the wave front, with the use of Eq. \eqref{eqn:G.1}.
We will return to the question of how $\sigma_\text{s}^\text{s}$ is affected by the uncertainty in $T^\text{s}$ in the following subsection and for the moment use $T^\text{s}$ as determined from the available data.
Figure \ref{fig:Ts} shows a plot of $\rho_\text{u}^{\text{s}}$ \textit{vs.} $\rho_\text{s}^{\text{s}}$.
The relationship is linear with slope $T^{\text{s}*}= 3.9$, which is between the upstream and downstream temperatures.
\begin{figure}[tbp]
\begin{center}
\includegraphics[trim=0 30 0 0, clip, width=1.0
\columnwidth]{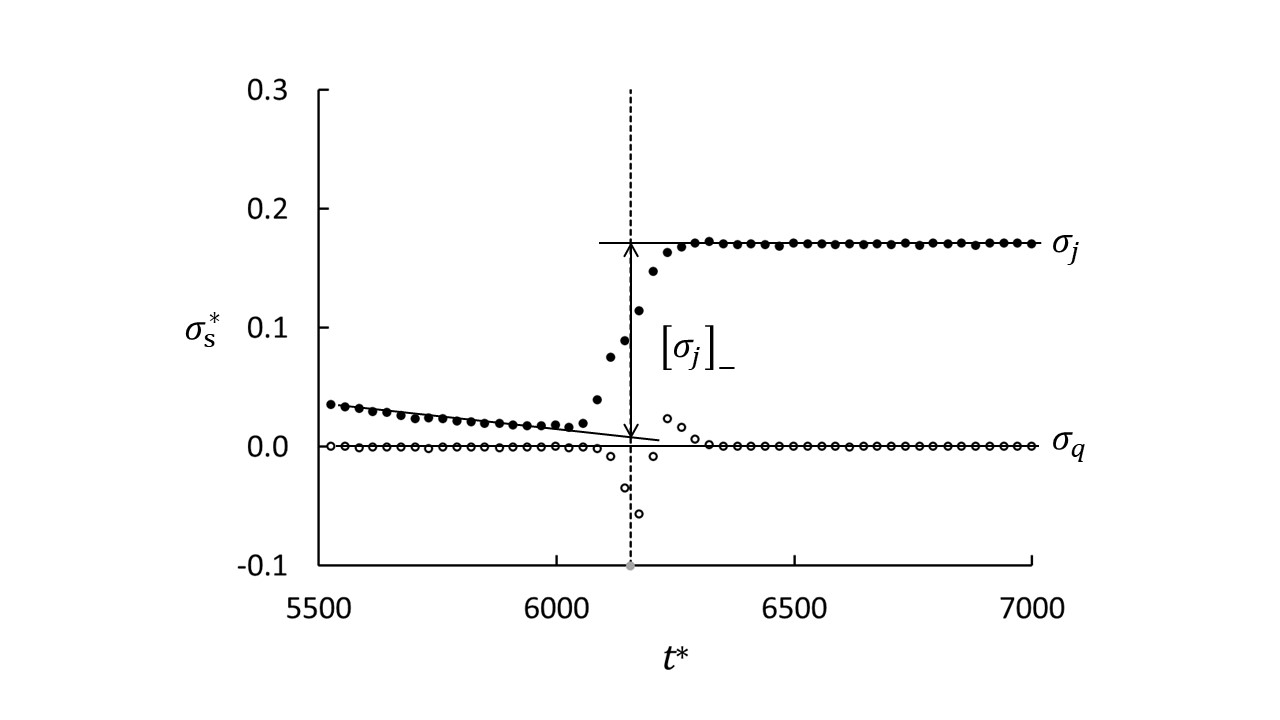}
\end{center}
\caption{Plots of $\sigma_q$ and $\sigma_j$ for the strong shock as determined from Eqs. \eqref{eqn:Na} and \eqref{eqn:Nb}, respectively, at $t^*=600$. The vertical dashed line shows the position of the Gibbs equimolar surface.}
\label{fig:sigmaqj}
\end{figure}

A plot of the local values of $\sigma_q$ and $\sigma_j$ (Eqs. \eqref{eqn:Na} and \eqref{eqn:Nb}) is shown in Fig. \ref{fig:sigmaqj} as function of $x^*$ for $t^*=600$.
The surface excess entropy production is the difference across the shock front as indicated by the double arrow in the figure.
The $\sigma_q$ is very small except in the shock front, where $\partial T/\partial x$ is large.
Ahead of the shock, $\sigma_q$ is zero because the system is at equilibrium there with $J_q^\prime=0$.
Behind the shock, $J_q^\prime \approx 0$, because $\partial T/\partial x$ is small there, cf. Fig. \ref{fig:temperature}.
The difference $[\sigma_q]_-$ of the extrapolated values is therefore practically zero.
The blip in the front region is due to the fact that $T^\text{d}<T^\text{s}<T^\text{u}$, but this is of no importance to the value of $[\sigma_q]_-$.

\begin{figure}[tbp]
\begin{center}
\includegraphics[trim=0 30 0 20, clip, width=0.9
\columnwidth]{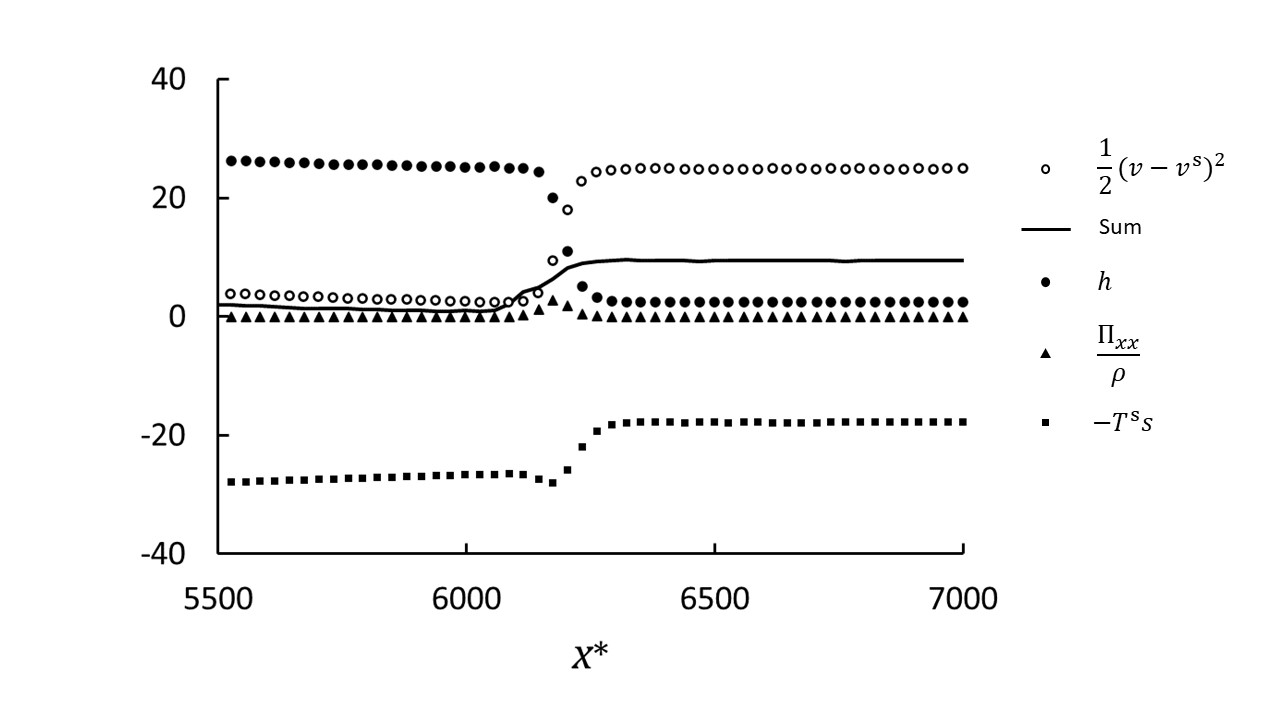}
\end{center}
\caption{The different terms in the parenthesis of Eq. \eqref{eqn:Nb} as function of $x$ at $t^*=600$. The line is the sum of the four terms.}
\label{fig:sigmaj}
\end{figure}

By comparison, $\sigma_j$ is everywhere large and $[\sigma_j]_-$ is significant.
Unlike $J_q^\prime$, the mass flux depends on the frame of reference, and $j$ in Eq. \eqref{eqn:Nb} is given with the surface as reference.
So is also the kinetic energy term in the parenthesis in Eq. \eqref{eqn:Nb}.
The local values of $\sigma_j$ must therefore not be confused with the local entropy production, it is the difference across the wave, indicated as the double arrow in Figure \ref{fig:sigmaqj}, which is relevant for $\sigma_\text{s}$.
The individual terms in the parenthesis of Eq. \eqref{eqn:Nb} are shown in Fig. \ref{fig:sigmaj} for the strong shock.
The viscous  pressure term varies little over the shock front and the difference between the extrapolated values is practically zero.
The kinetic energy term includes the center-of-mass velocity relative to the shock wave velocity.
This relative velocity is larger upstream than downstream, so the difference defined by the bracket is positive.
Both $h$ and $s$ increase when the shock wave passes.
The viscous term is negligible.
The mass flux is constant across the shock front because mass is conserved, and therefore equal to the upstream value, $j=-\rho v^\text{s}$.
In total, the term $[\sigma_j]_-$ is positive.
Hence, for the propagating shock examined in this work, the overall picture is that kinetic energy is converted to enthalpy.
A minor amount of the wave's energy produces entropy across the shock front, leading to a slow retardation of the wave.

\subsubsection{How sensitive are the results to uncertainties in the estimated surface temperature?}
\label{Ts}

In the GEM, the surface excess entropy production is determined according to Eqs. \eqref{eqn:N} - \eqref{eqn:Nb}.
All quantities in these equations, except the surface temperature $T^\text{s}$, are determined by extrapolating properties from the bulk phases adjacent to the wave front.
We established in Sections \ref{velocityprofiles} and \ref{pressureprofiles} that these properties are given by their equilibrium values in the present case.
The surface temperature is, however, given by Eq. \eqref{eqn:S.3}.
This equation involves the excess properties $\rho_\text{u}^{\text{s}}$ and $\rho_\text{s}^{\text{s}}$ determined from the entire profiles, including the non-equilibrium properties in the wave front, with the use of Eq. \eqref{eqn:G.1}.
We will therefore now estimate how much the uncertainty introduced by using the equilibrium entropy instead of the non-equilibrium entropy in Eq. \eqref{eqn:G.1} affects the value of the surface excess entropy production.

For shock waves near steady state and $J_q^{\prime\text{d}} \approx 0$, Eq. \eqref{eqn:sigmaapprox} holds.
In this case, $\sigma_\text{s}^\text{s}$ is independent of $T^\text{s}$.
Alternatively, we may consider the dominant contribution to $\sigma_\text{s}^\text{s}$, $\left [\sigma_j \right]_-$, and take the derivative with respect to $T^\text{s}$:
\begin{equation}
\frac{d \left [\sigma_j \right]_-}{d T^\text{s}} = \frac{j}{(T^\text{s})^2} \left [ h + \frac{\Pi_{xx}}{\rho} + \frac{1}{2} (v-v^\text{s})^2 \right ]_-
\label{eqn:sensitivity}
\end{equation}
The results in Section \ref{Gibbs} show that the bracket on the right hand side of Eq. \eqref{eqn:sensitivity} is small, which means that $\left [\sigma_j \right]_-$ is rather insensitive to errors in $T^\text{s}$.
As an example, inserting numerical values for $t^*=600$ shows that $[\sigma_j]_-$ changes by $\pm 0.1 \%$ for a $\pm 10 \%$ change in $T^\text{s}$.
Our conclusion is that the surface excess entropy production is very insensitive to the value of $T^\text{s}$ and that the Gibbs excess method is an accurate method in the present case.

\subsection{Entropy production and blast wave decay}
\label{comparison}

\begin{figure}[tbp]
\begin{center}
\includegraphics[trim=50 20 0 20, clip, width=0.9
\columnwidth]{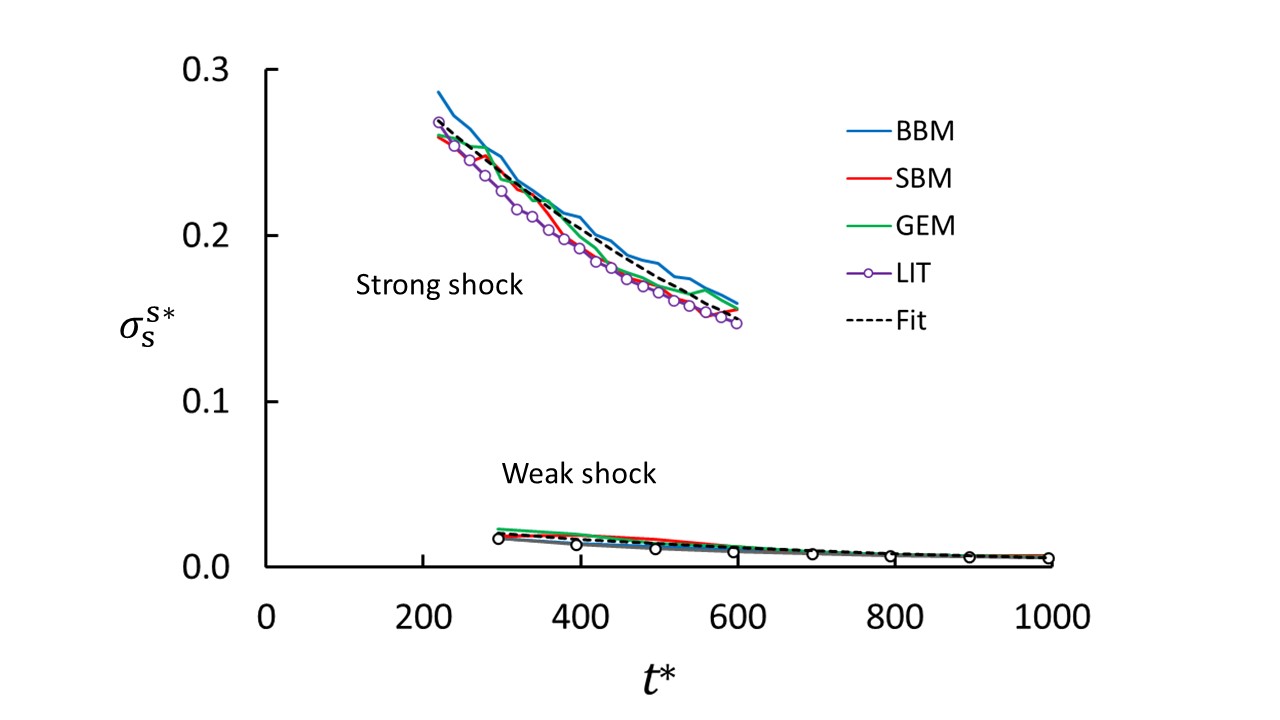}
\end{center}
\caption{Surface excess entropy production as function of time for the strong and weak shocks, Mach number $\approx 6$ and $\approx 2$, respectively. The three colored lines show data for the three methods used and the dashed lines are fitted exponential functions. The white dots are results obtained from the LIT method}.
\label{fig:finalresult}
\end{figure}

A comparison between the surface excess entropy production computed from the four methods employed in this work is shown in Figure \ref{fig:finalresult}.
The four methods are consistent for both Mach numbers, which adds confidence to the assumptions made.
The entropy production decreases with time as expected as the wave moves away from the blast, looses energy and slows down, cf. Figure \ref{fig:wavespeed}.

The four methods differ in the ways the sources are computed, but they give the same entropy production.
The time derivative of the entropy density and the space derivative of the entropy flux used in the BBM are both large in the front region and of opposite sign (\textit{cf} Figure \ref{fig:BBM2}), and the local entropy production is a small difference between relatively large numbers.
The BBM is therefore sensitive to errors in these quantities.
The SBM and the GEM depend on the time derivative of the excess surface entropy density, which varies little with time.
The main contributions in these methods are jumps in extrapolated quantities determined from bulk properties adjacent to the surface, which are robustly determined from the equilibrium equation of state.
The excess entropy production depends strongly on the Mach number, with approximately a factor 10 increase in the produced entropy when the Mach number increases from 2 to 6, or approximately the square of the ratio between the Mach numbers, $\left (M_\text{strong}/M_\text{weak} \right )^2$.
This difference is also reflected in the retardation of the shock wave, the strong shock slows down much faster than the weak shock.

A fitted exponential function, $\sigma_\text{s}^\text{s}=\sigma_0 \exp(-\alpha t)$, to the values for $\sigma_\text{s}^\text{s}$ gave the parameters $\sigma_0=0.034$ and $\alpha=0.0018$ for the weak shock and $\sigma_0=0.38$ and $\alpha=0.0015$ for the strong shock.
This means that the \textit{relative} decay is approximately the same for the two Mach numbers, but the intensities differ by a factor of approximately ten.

We have also included results from the assumption used by Velasco and Uribe, \textit{viz.} that the Gibbs equation is locally valid in each control volume (marked "LIT" in Figure \ref{fig:finalresult})\cite{velasco2019}.
These results are systematically lower the other three, albeit not by very much.
The difference may be an indication that although the local equilibrium assumption is good, it may lead to systematic errors in the computed entropy production.

\begin{figure}[tbp]
\begin{center}
\includegraphics[trim=90 30 0 20, clip, width=0.9
\columnwidth]{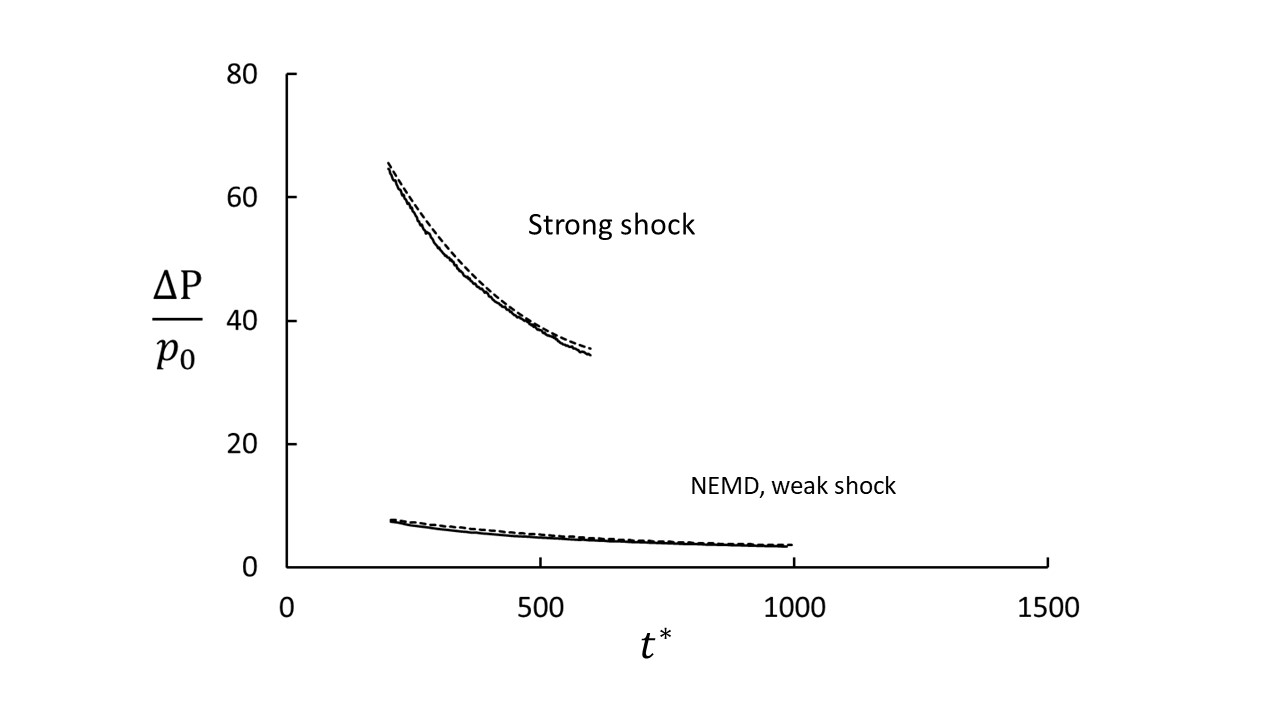}
\end{center}
\caption{Peak overpressure as function of time for the strong and weak shocks. The dashed lines represent the RH conditions (Eq. \eqref{eqn:peak}) and the solid lines are results from the NEMD simulations.}
\label{fig:peak}
\end{figure}

A key property in analyses of shock waves is the peak overpressure, \textit{i.e.} the maximum pressure in the shock wave minus the ambient pressure in front of the shock \cite{Friedlander1946}.
The peak overpressure, $\Delta \textsf{P}$, is given by Jones \cite{jones1968}
\begin{equation}
\Delta \textsf{P} = p_0 \left ( M^2 - 1 \right ) \frac{2 C_p/C_v}{1+C_p/C_v} 
\label{eqn:peak}
\end{equation}
where $p_0$ is the ambient pressure ahead of the shock, $C_p$ and $C_v$ are the heat capacities at constant pressure and volume, respectively, and $M$ is the Mach number.
The results from Eq. \eqref{eqn:peak} are compared with NEMD results in Figure \ref{fig:peak}.
For this comparison, the ambient pressure and the Mach number were taken from the NEMD simulations and the heat capacities were determined by separate MD simulations.
The agreement is good for both the weak and the strong shock.
Jones also gave a relation between the overpressure and the blast energy, which for a plane wave reads in our notation:
\begin{equation}
\Delta \textsf{P} = p_0 \left ( \frac{R_0}{\ell} \right )^m \frac{8 C_p/C_v}{9(1+C_p/C_v)} 
\label{eqn:m}
\end{equation}
where $R_0$ is a characteristic distance related to the blast energy \cite{jones1968}.
The exponent $m$ is equal to 1 in the limit of $M \rightarrow \infty$ and 1/2 for $M \rightarrow 1$ for a plane wave.
We found $m=0.87$ for the strong wave and $m=0.69$ for the weak wave.
Using our results together with the limiting values, we found the empirical relation $m \approx 1-0.82 (1/M) + 0.32 (1/M)^2$.
Finally, we note that the distance $R_0$ is proportional to the blast energy \cite{jones1968}.
Using $R_0$ as an adjustable parameter when fitting Eq. \eqref{eqn:m} to our NEMD data, we found the ratio $(R_0)_\text{strong}/(R_0)_\text{weak} = 21.9$, in excellent agreement with the ratio between the blast energies in the two cases, which was 22.0.

\section{Conclusions}
\label{conclusions}

In this work, we have applied non-equilibrium thermodynamics for surfaces \cite{kjelstrup2020} and analyzed the entropy production in two shock waves using four different methods.
We have developed a new tool, the "Gibbs excess method" (GEM) and compared it with three other methods.
In the "bulk balance method" (BBM), the entropy balance was integrated over the thickness of the shock wave.
The LIT method is based on the assumption of local equilibrium in the shock wave front, the local version of the Gibbs equation, and integration of the local entropy production over the shock thickness \cite{velasco2019}.
In the "surface balance method" (SBM), we used the concept of Gibbs equimolar surface combined with the entropy balance equation across the surface.
In the GEM, we took the SBM one step further by using the Gibbs equation for surfaces and derived new expressions for the surface excess entropy production, performed a detailed analysis of the energy conversions in the shock wave front, and found a very simple approximate expression for the entropy production.
The most significant difference between the four methods is that BBM and LIT assume local equilibrium everywhere in the fluid, including the shock front, whereas the SBM and the GEM use surface properties in equilibrium.
The SBM and GEM are therefore more robust and may be easier to apply.

Two plane blast waves were simulated with non-equilibrium molecular dynamics (NEMD) in a Lennard-Jones/spline system with 524,288 particles.
Prior to the blast, the system was equilibrated at $T^*=1.0$ and $\rho^*=0.01$.
The two shocks propagated at almost steady state with Mach numbers approximately 2 and 6.
We found the typical difference in the $x$- and $y$-components of the kinetic temperature, but based on analyses of the particle speeds, potential energy, and pressure, we concluded that the conditions for using non-equilibrium thermodynamics were well satisfied.

The four methods were based on different approximations and used the NEMD data in different ways, but the surface excess entropy productions were in excellent agreement.
We found a small deviation from local equilibrium in the front region of the strong shock, but this is of no importance in the GEM, which uses extrapolated equilibrium data from the adjacent bulk regions.
From this observation and verifications of some of the assumptions, we conclude that the results are reliable.
For the GEM, we found that the differences across the surface in the measurable heat flux and the viscous pressure gave negligible contributions to the entropy production.
This is in contrast to the LIT method, in which these are the only two sources to the entropy production.
The GEM provides new, detailed information about energy conversions in shock waves.
In short, most of the wave's kinetic energy is converted reversibly to enthalpy.
A smaller fraction of the waves total energy was dissipated, which led to a weak retardation of the wave.

In principle, the BBM makes no assumption of local equilibrium, but lacking data for the non-equilibrium entropy in the front region, we had to use equilibrium data here.
The SBM may be more robust that the BBM because the surface excess entropy density is almost constant with time and its time derivative is almost zero.
The LIT gives results with little statistical noise, but may suffer from the lack of local equilibrium in the front region.

As the waves were almost at steady states, the Rankine-Hugoniot (RH) conditions were found to describe the waves well.
The peak overpressure determined from the RH conditions agreed very well with the NEMD data. 
The shock-wave thickness was found to agree with theoretical estimates and experimental data \cite{bird1967} and simulations \cite{holian2010}.

The combination of different theories and NEMD data presented here gives new tools to study shock waves.
In particular, the GEM is a robust method that relies on equilibrium data adjacent to the shock wave front.
For waves close to steady state, a good appoximative value for the surface excess entropy production can be found in a very simple way as given by Eq. \eqref{eqn:sigmaapprox}, \textit{viz.} as the mass flux in the surface frame of reference times the difference in specific entropy across the surface.

\section*{Acknowledgements}

The authors are grateful to the Research Council of Norway, for the Center of Excellence Funding Scheme, project no 262644, PoreLab.
The NEMD simulations were performed on resources provided by UNINETT Sigma2 - the National Infrastructure for High Performance Computing and Data Storage in Norway and by Department of Chemistry at The Norwegian University of Science and Technology - NTNU.
We are grateful to Olav Galteland for providing data for the shear viscosity using LAMMPS with the SLLOD algorithm.\footnote{https://lammps.sandia.gov/doc/Howto\_viscosity.html}


\bibliographystyle{ieeetr}
\bibliography{shockwave1}

\begin{thebibliography}{10}

\bibitem{ning2015shock}
Y.-L. Ning and Y.-G. Zhou, ``Shock tubes and blast injury modeling,'' {\em
  Chinese Journal of Traumatology}, vol.~18, no.~4, pp.~187--193, 2015.

\bibitem{rankine1870}
W.~J.~M. Rankine {\em et~al.}, ``{XV}. {O}n the thermodynamic theory of waves
  of finite longitudinal disturbance,'' {\em Philosophical Transactions of the
  Royal Society of London}, vol.~160, pp.~277--288, 1870.

\bibitem{hugoniot1887}
H.~Hugoniot, ``Memoir on the propagation of movements in bodies, especially
  perfect gases (first part),'' {\em J. de l’Ecole Polytechnique}, vol.~57,
  pp.~3--97, 1887.

\bibitem{hirschfelder1954}
J.~O. Hirschfelder, C.~F. Curtiss, and R.~B. Bird, {\em Molecular Theory of
  Gases and Liquids}.
\newblock New York: John Wiley and Sons, 1954.

\bibitem{hoover2015}
W.~G. Hoover and C.~G. Hoover, {\em Simulation and Control of Chaotic
  Nonequilibrium Systems: With a Foreword by Julien Clinton Sprott}, vol.~27.
\newblock World Scientific Publishing Company, 2015.

\bibitem{uribe2011}
F.~J. Uribe, {\em The shock wave problem revisited: the Navier--Stokes
  equations and Brenner’s two velocity hydrodynamics}.
\newblock Springer, 2011.

\bibitem{jouguet1910}
E.~Jouguet, ``Sur la vitessee des ondes de choc et combustion,'' {\em Comptes
  rendus Académie des Sciences Paris}, vol.~149, p.~1361, 1910.

\bibitem{becker1922}
R.~Becker, ``Stosswelle und detonation,'' {\em Zeitschrift f{\"u}r Physik},
  vol.~8, no.~1, pp.~321--362, 1922.

\bibitem{bird1967}
G.~Bird, ``The velocity distribution function within a shock wave,'' {\em
  Journal of Fluid Mechanics}, vol.~30, no.~3, pp.~479--487, 1967.

\bibitem{morduchow1949}
M.~Morduchow and P.~A. Libby, ``On a complete solution of the one-dimensional
  flow equations of a viscous, heat-conducting, compressible gas,'' {\em
  Journal of the Aeronautical Sciences}, vol.~16, no.~11, pp.~674--684, 1949.

\bibitem{Friedlander1946}
F.~G. Friedlander, ``The diffraction of sound pulses {I}. {D}iffraction by a
  semi-infinite plane,'' {\em Proc. Roy. Soc. London}, vol.~A186, pp.~322 --
  344, 1946.

\bibitem{Taylor1950}
G.~I. Taylor, ``The formation of a blast wave by a very intense explosion {I}.
  {T}heoretical discussion,'' {\em Proc. R. Soc. Lond. A}, vol.~201, no.~1065,
  pp.~159--174, 1950.

\bibitem{Freiwald1972}
D.~Freiwald, ``Approximate blast wave theory and experimental data for shock
  trajectories in linear explosive-driven shock tubes,'' {\em Journal of
  Applied Physics}, vol.~43, no.~5, pp.~2224--2226, 1972.

\bibitem{margolin2020}
L.~G. Margolin, C.~S. Plesko, and J.~M. Reisner, ``A finite scale model for
  shock structure,'' {\em Physica D: Nonlinear Phenomena}, vol.~403, p.~132308,
  2020.

\bibitem{mott1951}
H.~M. Mott-Smith, ``The solution of the {B}oltzmann equation for a shock
  wave,'' {\em Physical Review}, vol.~82, no.~6, p.~885, 1951.

\bibitem{al1997generalized}
M.~Al-Ghoul and B.~C. Eu, ``Generalized hydrodynamics and shock waves,'' {\em
  Physical Review E}, vol.~56, no.~3, p.~2981, 1997.

\bibitem{taniguchi2014}
S.~Taniguchi, T.~Arima, T.~Ruggeri, and M.~Sugiyama, ``Thermodynamic theory of
  the shock wave structure in a rarefied polyatomic gas: Beyond the
  {B}ethe-{T}eller theory,'' {\em Physical Review E}, vol.~89, no.~1,
  p.~013025, 2014.

\bibitem{arima2012}
T.~Arima, S.~Taniguchi, T.~Ruggeri, and M.~Sugiyama, ``Extended thermodynamics
  of dense gases,'' {\em Continuum Mechanics and Thermodynamics}, vol.~24,
  no.~4-6, pp.~271--292, 2012.

\bibitem{garcia2008}
L.~S. Garc{\'\i}a-Col{\'\i}n, R.~M. Velasco, and F.~J. Uribe, ``Beyond the
  {N}avier-{S}tokes equations: {B}urnett hydrodynamics,'' {\em Physics
  Reports}, vol.~465, no.~4, pp.~149--189, 2008.

\bibitem{klimenko1978}
V.~Y. Klimenko and A.~N. Dremin, ``The structure of the shock wave front in
  liquids,'' in {\em Detonation, critical phenomena, physico-chemical
  transformation in shock waves (ed. 0. N. Breusov)}, pp.~79--84,
  Chernogolovka: Nauka, 1978.

\bibitem{hoover1979}
W.~G. Hoover, ``Structure of a shock-wave front in a liquid,'' {\em Physical
  Review Letters}, vol.~42, no.~23, pp.~1531--1534, 1979.

\bibitem{holian1980}
B.~L. Holian, W.~G. Hoover, B.~Moran, and G.~K. Straub, ``Shock-wave structure
  via nonequilihrium molecular dynamics and {N}avier-{S}tokes continuum
  mechanics,'' {\em Physical Review A}, vol.~22, no.~6, pp.~2798--2808, 1980.

\bibitem{salomons1992}
E.~Salomons and M.~Mareschal, ``Usefulness of the {B}urnett description of
  strong shock waves,'' {\em Physical Review Letters}, vol.~69, no.~2,
  pp.~269--272, 1992.

\bibitem{holian1993}
B.~L. Holian, C.~Patterson, M.~Mareschal, and E.~Salomons, ``Modeling shock
  waves in an ideal gas: Going beyond the {N}avier-{S}tokes level,'' {\em
  Physical review E}, vol.~47, no.~1, p.~R24, 1993.

\bibitem{holian1995}
B.~Holian, ``Atomistic computer simulations of shock waves,'' {\em Shock
  waves}, vol.~5, no.~3, pp.~149--157, 1995.

\bibitem{uribe2018}
F.~Uribe and R.~Velasco, ``Shock-wave structure based on the
  {N}avier-{S}tokes-{F}ourier equations,'' {\em Physical Review E}, vol.~97,
  no.~4, p.~043117, 2018.

\bibitem{brinkley1947}
S.~R. Brinkley~Jr and J.~G. Kirkwood, ``Theory of the propagation of shock
  waves,'' {\em Physical Review}, vol.~71, no.~9, p.~606, 1947.

\bibitem{tolman1948}
R.~C. Tolman and P.~C. Fine, ``On the irreversible production of entropy,''
  {\em Reviews of Modern Physics}, vol.~20, no.~1, p.~51, 1948.

\bibitem{holian2010}
B.~L. Holian and M.~Mareschal, ``Heat-flow equation motivated by the ideal-gas
  shock wave,'' {\em Physical Review E}, vol.~82, no.~2, p.~026707, 2010.

\bibitem{velasco2019}
R.~Velasco and F.~Uribe, ``Shock-wave structure according to a linear
  irreversible thermodynamic model,'' {\em Physical Review E}, vol.~99, no.~2,
  p.~023114, 2019.

\bibitem{kinney2013}
G.~F. Kinney and K.~J. Graham, {\em Explosive shocks in air}.
\newblock Springer Science \& Business Media, 2013.

\bibitem{Zhao2017}
S.~Zhao, B.~Kad, C.~E. Wehrenberg, B.~A. Remington, E.~N. Hahn, K.~L. More, and
  M.~A. Meyers, ``Generating gradient germanium nanostructures by shock-induced
  amorphization and crystallization,'' {\em Proceedings of the National Academy
  of Sciences}, vol.~114, no.~37, pp.~9791--9796, 2017.

\bibitem{Pecha2000}
R.~Pecha and B.~Gompf, ``Microimplosions: Cavitation collapse and shock wave
  emission on a nanosecond time scale,'' {\em Phys. Rev. Lett.}, vol.~84,
  pp.~1328--1330, Feb 2000.

\bibitem{kjelstrup2020}
S.~Kjelstrup and D.~Bedeaux, {\em Non-equilibrium thermodynamics of
  heterogeneous systems}, vol.~20.
\newblock World Scientific, 2020.

\bibitem{hafskjold2020}
B.~Hafskjold, D.~Bedeaux, S.~Kjelstrup, and {\O}.~Wilhelmsen, ``Nonequilibrium
  thermodynamics of surfaces captures the energy conversions in a shock wave,''
  {\em Chemical Physics Letters: X}, vol.~7, p.~100054, 2020.

\bibitem{deGroot1962}
S.~R. de~Groot and P.~Mazur, {\em Non-Equilibrium Thermodynamics}.
\newblock Amsterdam: North-Holland, 1962.

\bibitem{Bedeaux1976}
D.~Bedeaux, A.~M. Albano, and P.~Mazur, ``Boundary conditions and
  non-equilibrium thermodynamics,'' {\em Physica A}, vol.~82, pp.~438--462,
  1976.

\bibitem{Albano1987}
A.~M. Albano and D.~Bedeaux, ``Non-equilibrium electro-thermodynamics of
  polarizable multicomponent fluids with an interface,'' {\em Physica A},
  vol.~147, no.~1-2, pp.~407--435, 1987.

\bibitem{Gibbs1961}
J.~W. Gibbs, {\em The Scientific Papers, Vol I: Thermodynamics}.
\newblock New York: Dover, 1961.

\bibitem{holian1983}
B.~L. Holian and D.~J. Evans, ``Shear viscosities away from the melting line: A
  comparison of equilibrium and nonequilibrium molecular dynamics,'' {\em The
  Journal of chemical physics}, vol.~78, no.~8, pp.~5147--5150, 1983.

\bibitem{hafskjold2019}
B.~Hafskjold, K.~P. Travis, A.~B. Hass, M.~Hammer, A.~Aasen, and
  {\O}.~Wilhelmsen, ``Thermodynamic properties of the 3d
  {L}ennard-{J}ones/spline model,'' {\em Molecular Physics}, vol.~117,
  no.~23-24, pp.~3754--3769, 2019.

\bibitem{hafskjold1993}
B.~Hafskjold, T.~Ikeshoji, and S.~K. Ratkje, ``On the molecular mechanism of
  thermal diffusion in liquids,'' {\em Molecular Physics}, vol.~80, no.~6,
  pp.~1389--1412, 1993.

\bibitem{walton1983}
J.~Walton, D.~Tildesley, J.~Rowlinson, and J.~Henderson, ``The pressure tensor
  at the planar surface of a liquid,'' {\em Molecular physics}, vol.~48, no.~6,
  pp.~1357--1368, 1983.

\bibitem{ikeshoji2003}
T.~Ikeshoji, B.~Hafskjold, and H.~Furuholt, ``Molecular-level calculation
  scheme for pressure in inhomogeneous systems of flat and spherical layers,''
  {\em Molecular Simulation}, vol.~29, no.~2, pp.~101 -- 109, 2003.

\bibitem{irving1950}
J.~Irving and J.~G. Kirkwood, ``The statistical mechanical theory of transport
  processes. {IV}. {T}he equations of hydrodynamics,'' {\em The Journal of
  Chemical Physics}, vol.~18, no.~6, pp.~817--829, 1950.

\bibitem{evans1990}
D.~Evans and G.~Morriss, {\em Statistical Mechanics of Nonequilibrium Liquids
  (Academic, London, 1990)}.
\newblock London: Academic, 1990.

\bibitem{hoover2014}
W.~G. Hoover, C.~G. Hoover, and K.~P. Travis, ``Shock-wave compression and
  {J}oule-{T}homson expansion,'' {\em Phys. Rev. Lett.}, vol.~112, p.~144504,
  Apr 2014.

\bibitem{hoover2016}
C.~G. Hoover and W.~G. Hoover, ``Yokohama to {R}uby valley: Around the world in
  80 years. ii,'' {\em arXiv preprint arXiv:1606.03183}, 2016.

\bibitem{meier2005}
K.~Meier, A.~Laesecke, and S.~Kabelac, ``Transport coefficients of the
  {L}ennard-{J}ones model fluid. {III}. {B}ulk viscosity,'' {\em The Journal of
  Chemical Physics}, vol.~122, no.~1, p.~014513, 2005.

\bibitem{plimpton1995}
S.~Plimpton, ``Fast parallel algorithms for short-range molecular dynamics,''
  {\em Journal of computational physics}, vol.~117, no.~1, pp.~1--19, 1995.

\bibitem{jones1968}
D.~L. Jones, ``Intermediate strength blast wave,'' {\em The Physics of Fluids},
  vol.~11, no.~8, pp.~1664--1667, 1968.

\end{thebibliography}
\newpage

\section*{Supplementary material}
\label{appendix}

\begin{table}[!ht]
\centering
\caption{\textbf{Symbol list}}
\vspace{0.2 cm}
\begin{tabular}{c c l}\hline
Symbol & Definition & Meaning \\
\hline \\
\vspace{0.1 cm}
$C_p, C_v$ &  &Heat capacity at constant pressure, constant volume \\
$\delta(x)$ &  &Dirac delta function \\
$\varepsilon$ &  &Lennard-Jones parameter \\
$\mathbf{f}_{ij}$ &  &Force on particle $i$ due to $j$ \\
$\phi_i$ &  &Potential energy of particle $i$ \\
$\eta_\text{S}, \eta_\text{B}$ &  &Shear viscosity, bulk viscosity \\
$\gamma$ & $(\partial U^\text{s}/\partial \Omega)_{\{ S^\text{s},N^\text{s}\} }$ &Surface property \\
$h$ & &Specific enthalpy \\
$j$ & &Mass flux in the surface frame of reference \\
$J$ &  &Flux \\
$k_\text{B}$ &  &Boltzmann's constant \\
$L,L_{aa}$ &  &Onsager coefficient (of property "$a$") \\
$L_q$ &  &Length of MD cell in $q$-direction \\
$\ell$ &  &Equimolar surface position \\
$\lambda$ &  &Thermal conductivity or mean free path \\
$m$ &  &Particle mass \\
$\mu$ &  &Specific Gibbs energy \\
$N$ &  &Number of particles \\
$p$ &  &Thermodynamic equilibrium pressure \\
$\Delta \textsf{P}$ &  &Shock overpressure \\
\vspace{0.1 cm}
$\textsf{P}_{qq}$ & $p+\Pi_{qq}$   &$qq$-component of pressure tensor \\
$\Pi_{qq}$ & $-\left ( \frac{4}{3}\eta_\text{S}+\eta_\text{B} \right ) \frac{\partial v}{\partial x}$   &$x$-component of viscous pressure  \\
$\rho$ & &Density \\
$\sigma$ & &Lennard-Jones parameter \\
$\sigma_\text{s}$ & &Entropy production rate \\
$r$ & &Interparticle distance \\
$r_s, r_c$ & &Lennard-Jones parameters \\
$s$ & &Specific entropy \\
$t$ & &Time \\
$T$ & &Thermodynamic temperature \\
$\textsf{T}, \textsf{T}_{qq}$ & &Kinetic temperature, $qq$-component of $\textsf{T}$ \\
$\Theta(x)$ & &Heaviside step function \\
$u$ & &Specific internal energy \\
$V$ & &Volume \\
$\mathbf{v}_i$ & &Velocity of particle $i$ \\
$v$ & &Streaming velocity in $x$-direction \\
$v^\text{s}$ & $\partial \ell /\partial t$ &Surface speed \\
$x$ & &Position in $x$-direction \\
\hline
\end{tabular}
\label{symbols}    
\end{table}

\begin{table}[!h]
\centering
\caption{\textbf{Subscripts}}
\vspace{0.2 cm}
\begin{tabular}{c l}\hline
Symbol & Meaning \\
\hline \\
a &Property "a" \\
CV &Control volume, layer \\
e &Total energy \\
$i,j$ &Particle labels \\
k &Kinetic energy \\
$q$ &Heat \\
$qq$ &$q$-component, $q=x,y,z$ \\
s &Entropy \\
$T$ &Constant temperature \\
u &Potential energy \\
$x$ &$x$-direction \\
\hline
\end{tabular}
\label{subscripts}    
\end{table}

\begin{table}[h!]
\centering
\caption{\textbf{Superscripts}}
\vspace{0.2 cm}
\begin{tabular}{c l}\hline
Symbol & Meaning \\
\hline \\
\vspace{0.1 cm}
d &Downstream, behind the shock wave \\
s &Surface \\
u &Upstream ahead of the shock wave\\
\hline
\end{tabular}
\label{superscripts}    
\end{table}

\newpage
\begin{table}[!ht]
\centering
\caption{\textbf{Definitions of reduced variables}}
\vspace{0.4 cm}
\begin{tabular}{c c l}\hline
Symbol & Definition & Meaning \\
\hline \\
\vspace{0.2 cm}
$j^*$ & $j\frac{\sigma^3}{\left (m\varepsilon \right )^{1/2}}$ & Mass flux \\
\vspace{0.2 cm}
$J_q^*$ & $J_q\frac{\sigma^3}{\varepsilon} \left (\frac{m}{\varepsilon}\right )^{1/2}$ & Heat flux \\
\vspace{0.2 cm}
$J_s^*$ & $J_s\frac{\sigma^3}{k_\text{B}} \left (\frac{m}{\varepsilon}\right )^{1/2}$ & Entropy flux \\
\vspace{0.2 cm}
$\ell^*$ & $\ell/\sigma$ & Shock-wave position \\
\vspace{0.2 cm}
$M$ & $v^\text{s}/v_\text{sound}$ & Mach number, wave speed divided by speed of sound \\
\vspace{0.2 cm}
$Pr$ & $\frac{C_p\mu}{\lambda}$ & Prandtl number \\
\vspace{0.2 cm}
$p^*,\textsf{P}^*,\Pi^*$ & $\frac{p\sigma^3}{\varepsilon},\frac{\textsf{P}\sigma^3}{\varepsilon},\frac{\Pi \sigma^3}{\varepsilon}$ & Pressure, pressure tensor, viscous pressure \\
\vspace{0.2 cm}
$s^*$ & $s\frac{m}{k_\text{B}}$ & Specific entropy \\
\vspace{0.2 cm}
$t^*$ & $t \frac{1}{\sigma}\left (\frac{\varepsilon}{m}\right )^{1/2}$ & Time \\
\vspace{0.2 cm}
$T^*$ & $\frac{k_\text{B}T}{\varepsilon}$ & Temperature \\
\vspace{0.2 cm}
$u^*,h^*$ & $u\frac{m}{ \varepsilon},h\frac{m}{\varepsilon}$ & Specific internal energy, specific enthalpy \\
\vspace{0.2 cm}
$v^*$ & $v \left (\frac{m}{\varepsilon}\right )^{1/2}$ & Velocity \\
\vspace{0.2 cm}
$x^*$ & $x/\sigma$ & $x$-coordinate \\
\vspace{0.2 cm}
$\eta^*$ & $\eta \frac{\sigma^2}{\left (m \varepsilon\right )^{1/2}}$ & Viscosity \\
\vspace{0.2 cm}
$\lambda^*$ & $\lambda\frac{\sigma^2}{k_\text{B}} \left (\frac{m}{\varepsilon}\right )^{1/2}$ & Thermal conductivity \\
\vspace{0.2 cm}
$\lambda^*$ & $\lambda/\sigma$ & Molecular mean free path \\
\vspace{0.2 cm}
$\rho^*$ & $\rho \sigma^3$ & Number density, mass density \\
\vspace{0.2 cm}
$\rho^{\text{s}*}$ & $\rho^{\text{s}}\sigma^2$ & Surface excess number density, surface excess mass density \\
\vspace{0.2 cm}
$\rho_\text{s}^*$ & $\rho_\text{s} \frac{\sigma^3}{k_\text{B}}$ & Entropy density \\
\vspace{0.2 cm}
$\rho_\text{s}^{\text{s}*}$ & $\rho_\text{s}^{\text{s}}\frac{\sigma^2}{k_\text{B}}$ & Surface excess entropy density \\
\vspace{0.2 cm}
$\sigma^*$ & $\sigma\frac{\sigma^3}{k_\text{B}} \left (\frac{m}{\varepsilon}\right )^{1/2}$ & Entropy production \\
\hline
\end{tabular}
\label{reduced variables}    
\end{table}

\clearpage
\renewcommand\thesection{\Alph{section}}
\setcounter{section}{0}
\setcounter{equation}{0}
\numberwithin{equation}{section}
\numberwithin{table}{section}

\section{Equation of state for the Lennard-Jones/spline gas}

We consider here a one-component Lennard-Jones/spline (LJ/s) fluid at low density with the purpose to find an expression for its entropy and internal energy. The entropy, $S$, can be derived from the Helmholtz energy, $A$, as
\begin{equation}
S=-\left ( \frac{\partial A}{\partial T} \right )_V
\end{equation}
where $T$ is temperature and $V$ is volume. The Helmholtz energy can be found by integrating the $P,V$ equation of state at some constant temperature $T$
\begin{equation}
A=-\int P(V,T) dV
\end{equation}
Likewise, the internal energy can be found from 
\begin{equation}
U=\left ( \frac{\partial A/T}{\partial 1/T} \right )_V
\end{equation}
At low density, the virial expansion is a good representation of the equation of state, and we have used the expansion presented by Hafskjold et al. \cite{hafskjold2019},
\begin{equation}
\frac{P}{nk_\text{B} T}= 1+\sum_{k=2}^m B_k(T) n^{k-1}
\end{equation}
where $n=N/V$ is the number density of particles and $k_\text{B}$ is Boltzmann's constant. Using the polynomial fit in inverse temperature given in ref \cite{hafskjold2019}, the first three virial coefficients are 
\begin{equation}
B_k(T)=\sum_{l=0}^m a_{k,l} T^{-l}
\label{eqn:virialcoeff}
\end{equation}
with the coefficients given in Table \ref{eqn:virial_fit_coeff}. (Ref to the LJ/spline paper.)
\begin{table}[h!]
\centering
\caption{Fitted coefficients $a_{k,n}$ of the inverse temperature relation, Eq. \eqref{eqn:virialcoeff} for the virial coefficients $B_2,B_3$, and $B_4$. The uncertainties represent 95\% confidence intervals.}
\vspace{0.4 cm}
\begin{tabular}{c c c c}\hline
$l$ & $k=2$ & $k=3$ & $k=4$ \\
\hline
$0$ & $1.345\pm 0.007$  & $3.76 \pm 0.02$  & $-1.376 \pm 0.007$ \\
$1$ & $-1.336 \pm 0.007$  & $  -20.9 \pm 0.1 $ & $ 40.5 \pm 0.2 $ \\
$2$ & $-3.85 \pm 0.02$ & $ 64.0 \pm 0.3 $ & $ -260 \pm 1 $ \\
$3$ & $1.295 \pm 0.006$ &  $ -90.2 \pm 0.5 $ & $ 936 \pm 5 $ \\
$4$ & $-0.416 \pm 0.002$ &  $ 66.8 \pm 0.3 $ & $ -2069 \pm 10 $  \\
$5$ & $-$ & $ -20.1 \pm 0.1 $ & $ 2789 \pm 14 $ \\
$6$ & $-$ & $-$ & $ -2240 \pm 11 $ \\
$7$ & $-$ & $-$ & $ 1010 \pm 5 $ \\
$8$ & $-$ & $-$ & $ -200 \pm 1 $ \\\hline
\end{tabular}
\label{eqn:virial_fit_coeff}    
\end{table}

The corresponding expression for the Helmholtz free energy per particle, $a$, is:
\begin{equation}
\beta a =\ln (n\Lambda^3) -1 +\sum_{k=2}^\infty \frac{n^{k-1}}{k-1}B_k(\beta)
\label{eqn:freeenergy2}
\end{equation}
where $\beta=1/k_\text{B}T$ and $\Lambda$ is the thermal de Broglie wave length.
This gives for the entropy per particle, $s$, 
\begin{equation}
s=-\left ( \frac{\partial a}{\partial T} \right )_n = s^\text{ig} + s^\text{ex}
\end{equation}
where
\begin{equation}
s^\text{ig}=constant + k_\text{B} \left( \frac{3}{2}\ln T - \ln n \right )
\label{eqn:idealentropy}
\end{equation}
is the Sackur-Tetrode expression for the ideal-gas entropy, and 
\begin{equation}
s^\text{ex}=k_\text{B}\sum_{k=2}^\infty \sum_{l=0}^m \frac{l-1}{k-1}  a_{k,l} n^{k-1} T^{-l}
\label{eqn:excessentropy}
\end{equation}
is the virial expansion for the non-ideal contribution.
In Eq. \eqref{eqn:idealentropy}, $constant$ includes the terms in $s^\text{ig}$ that do not depend on either $n$ or $T$.

Similarly, we get for the internal energy per particle, $u$,
\begin{equation}
u=\left ( \frac{\partial (\beta a)}{\partial \beta} \right )_n=u^\text{ig} + u^\text{ex}
\end{equation}
where
\begin{equation}
u^\text{ig} = \frac{3}{2} k_\text{B}T
\end{equation}
and 
\begin{equation}
u^\text{ex} = \sum_{k=2}^\infty \sum_{l=1}^m \frac{l}{k-1}  a_{k,l} n^{k-1} T^{-l}
\end{equation}

Note that the corresponding densities are obtained as $\rho_\text{s}=n s$ and $\rho_\text{u}=n u$.

\newpage
\newpage

\end{document}